\documentclass[11pt,letterpaper]{article}%
\usepackage{amsmath}
\usepackage{epsfig}
\usepackage{graphicx}%
\usepackage{amsfonts}%
\usepackage{amssymb}

\begin{document}
%
\begin{titlepage}
\begin{flushright}
UCL-IPT-03-15
\end{flushright}
\vspace*{30mm}
\begin{center}
\huge
{Searching for dominant rescattering sources\\in B to two pseudoscalar decays}
\end{center}
\vspace*{10mm}
\begin{center}
\Large{Christopher Smith\footnote{smith@fyma.ucl.ac.be}}
\end{center}
\vspace*{5mm}
\begin{center}
Institut de Physique Th\'{e}orique, Universit\'{e}
catholique de Louvain\\
Chemin du Cyclotron, 2, B-1348, Louvain-la-Neuve, Belgium
\end{center}
\vspace*{5mm}
\begin{center}
September 7, 2003
\end{center}
\vspace*{10mm}
\begin{abstract}
Various rescattering sources are analyzed in the context of the SU(3) flavor symmetry.
In particular, the possibility to account for intermediate charm at the hadronic level in B to PP is
thoroughly investigated. Then, the rescattering sources are compared in light of
recent B to two charmless pseudoscalar decay measurements, with emphasis
on the size of strong phases and on patterns of direct CP-asymmetries.
\end{abstract}
\end{titlepage}%

\newpage

\section{Introduction}

Studies of CP violation in the $B$ system offer a unique window into the
intimate structure of the Standard Model, and possibly about New Physics. The
dedicated experiments at Cornell, KEK and SLAC have begun to feed theorists
with precise data to test the current understanding of CP violation, as well
as alternate theories.

Recently, a number of the rare hadronic decays $B$ to two charmless
pseudoscalars have been measured (see [1-6]). These modes are interesting to
constrain the weak angle $\gamma$ of the unitary triangle. However, for this
program to be achieved, theoretical control is needed over the strong final
state interactions (FSI). In the present work, this problem is addressed
phenomenologically, at the hadronic level (for other approaches, see for
example \cite{OtherWorks}). The basic tools are flavor symmetries like isospin
or SU(3) (see for example \cite{su(3)}, \cite{Gronau}, \cite{MyPaper}) and
Watson's theorem, or elasticity, for strong FSI (see \cite{MyPaper},
\cite{Watson}, \cite{PREVIOUS}).

Up to now, a serious shortcoming of elasticity-based approaches was their
inability to account for intermediate charm. In short-distance analyses, quark
diagrams involving charm quarks do have non-negligible imaginary parts
\cite{GerardHou}. This led to the common belief that rescattering cannot be
elastic. One purpose of the present paper is to show that it is possible
within the hadronic framework of Watson's theorem elasticity to account for
intermediate charm. Some results in that direction were already presented
earlier \cite{PREVIOUS}. Here, we will revert to the language of the usual
quark diagram topologies. The theoretical foundations of our parametrization,
which are articulated around a factorization property for FSI, are reviewed in
the first part of the paper.

Another problem of elasticity-based approaches is that large strong phases are
apparently required by current experimental data for $B$ to two charmless
pseudoscalars (see for example \cite{Hou}). This is in disagreement with Regge
computations \cite{Regge}, with the expectation for two particles flying apart
with high momentum \cite{MBInfinite} and with QCD-factorization predictions
\cite{QCDFa}. In the second part of the paper, we will show that when
intermediate charm is accounted for, the strong phases tend to be much smaller.

Finally, the last goal of the paper is to parametrize and quantify the
corrections to the patterns for $B\rightarrow\pi K$ direct CP-asymmetries
presented in \cite{PREVIOUS}. These patterns were designed to discriminate
among dominant rescattering sources, and we will here show that this
capability is not altered.

\section{Theoretical Framework}

The starting point of our analysis is the ''generalized Watson Theorem'' (for
details, see \cite{MyPaper}, \cite{Watson}, \cite{PREVIOUS}), which permits
the factorization of weak decay amplitudes into a factor invariant under CP
and a factor that gets complex conjugated under CP%
\begin{subequations}
\begin{align}
W  &  =\sqrt{S}W_{b}\label{Eq1}\\
CP\left(  W\right)   &  =\sqrt{S}W_{b}^{\ast}%
\end{align}
where $W$ is a vector containing the full decay amplitudes. Consequently, the
real (up to CKM factors) $W_{b}$ is identified as the \textit{bare decay
amplitude} (i.e. before FSI) denoted as
\end{subequations}
\begin{equation}
W=\left(
\begin{array}
[c]{c}%
B\rightarrow X_{1}\\
\vdots
\end{array}
\right)  ,\;\;W_{b}=\left(
\begin{array}
[c]{c}%
B\rightarrow\left\{  X_{1}\right\} \\
\vdots
\end{array}
\right)  \label{Eq2}%
\end{equation}
The unitary \textit{rescattering matrix} $\sqrt{S}$ contains all the
CP-conserving strong phases and describes final state interactions (FSI). In
general, one defines the rescattering \textit{eigenchannels} $C_{i}$
\begin{equation}
\left(
\begin{array}
[c]{c}%
C_{1}\\
\vdots
\end{array}
\right)  =O_{C}\left(
\begin{array}
[c]{c}%
\left\{  X_{1}\right\} \\
\vdots
\end{array}
\right)  \label{Eq3}%
\end{equation}
as the basis in which $\sqrt{S}$ is diagonal%
\begin{equation}
S_{diag}=O_{C}.S.O_{C}^{t}=\left(
\begin{array}
[c]{ccc}%
e^{2i\delta_{C_{1}}} & 0 & \cdots\\
0 & e^{2i\delta_{C_{2}}} & \cdots\\
\vdots & \vdots & \ddots
\end{array}
\right)  \label{Eq4}%
\end{equation}
with strong phases (or rescattering \textit{eigenphases}) $\delta_{C_{i}}$ as
diagonal elements. \textit{Elasticity} (sometimes referred to as
''quasi-elasticity'') is defined as the conservation of the total decay
probability, which follows from the unitarity of $\sqrt{S}$
\begin{equation}
\left\|  W\right\|  ^{2}=\left\|  \sqrt{S}W_{b}\right\|  ^{2}=\left\|
W_{b}\right\|  ^{2} \label{Eq4a}%
\end{equation}
Assuming that CP and CPT hold for the strong interactions generating $\sqrt
{S}$ implies that $O_{C}$ is orthogonal (up to some phase conventions) since
then $\sqrt{S}$ is unitary and symmetric. The angles parametrizing $O_{C}$ are
then called \textit{mixing angles} since they govern the FSI mixings among
decay channels.

In principle, $\sqrt{S}$ should be a $n\times n$ matrix with $n$ the number of
$B$-meson decay channels which can be connected by strong interactions. In
practice, however, we expect that to a good approximation this matrix is
bloc-diagonal. Unfortunately, it is not known at present how large the extent
of each bloc should be to capture the essential rescattering physics. In the
present work, we assume that two-body $\rightleftharpoons$ $n$-body
rescatterings are negligible (as could be justified from $1/N$ arguments
\cite{1N}), or at least suppressed by large cancellations.

The most restrictive (though non-trivial) approximation is that of
SU(2)-elasticity: each bloc of $\sqrt{S}$ corresponds to an isospin multiplet,
which means that only rescatterings like $\pi\pi\rightleftharpoons\pi\pi$ or
$K\bar{K}\rightleftharpoons K\bar{K}$ (see \cite{PREVIOUS}) are allowed. A
more flexible approach is to consider SU(3)-elasticity, to open rescatterings
between different iso-multiplets like $\pi\pi\rightleftharpoons K\bar{K}$ or
$\pi K\rightleftharpoons\eta_{8}K$. In an attempt to capture all the relevant
physics, rescattering channels like $D\bar{D}$ $\rightleftharpoons\pi\pi$ or
$D_{s}\bar{D}$ $\rightleftharpoons\pi K$ should also be opened to account for
intermediate charm. However, sizeable $D\bar{D}$ $\rightleftharpoons PP$
rescatterings like those implied by SU(4)-elasticity would average decay
amplitudes and produce $Br\left(  B\rightarrow D\bar{D}\right)  \sim Br\left(
B\rightarrow PP\right)  $, in clear disagreement with experiment
\begin{equation}
Br^{\exp}\left(  B\rightarrow D_{s}\bar{D}\right)  \sim10^{-2}>>Br^{\exp
}\left(  B\rightarrow K\pi\right)  \sim10^{-5} \label{Eq4b}%
\end{equation}
Intermediate charm rescattering effects are thus quite small, and can be
accounted for as distortions of SU(3)-elasticity. We call \underline{enlarged}
SU(3)-elasticity this parametrization of FSI, since the SU(3) flavor symmetry
remains exact. Conversely, even if all the rescatterings (both two-body and
multi-body) are small, the $D\bar{D}$ $\rightleftharpoons PP$ ones may still
lead to sizeable effects because of Eq.(\ref{Eq4b}), so it is desirable to
have an adequate formalism at hand.

By definition, bare amplitudes $W_{b}$ are real (except for CKM matrix
elements) since they do not contain any FSI effects (see Eqs.(1)). To
parametrize them, and to make contact with the short-distance weak decay of
the $b$ quark, it is convenient to introduce quark diagrams which account for
the possible flavor flows (see, for example, \cite{su(3)}, \cite{Gronau},
\cite{MyPaper}). To first order in the weak interactions, and to all orders in
the strong interactions (except FSI), the six topologies needed for
$B\rightarrow\left\{  PP\right\}  $, $P=\pi,K,\eta_{8}$ and $B\rightarrow
\left\{  D\bar{D}\right\}  $, $D=D,D_{s}$ are shown in Fig.1.%
\[%
\begin{tabular}
[c]{c}%
\begin{tabular}
[c]{ccc}%
{\includegraphics[
height=0.9115in,
width=1.1026in
]%
{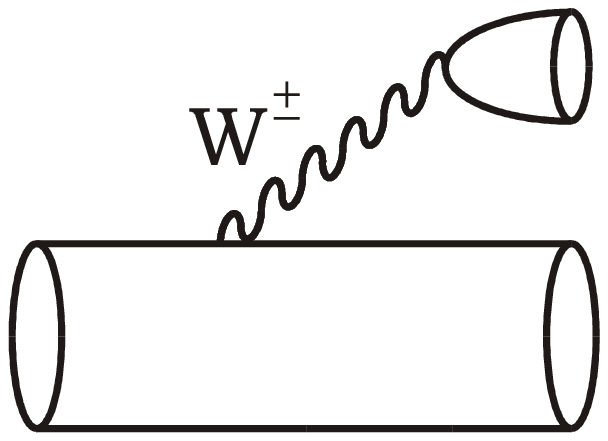}%
}%
&
{\includegraphics[
height=0.9115in,
width=1.0101in
]%
{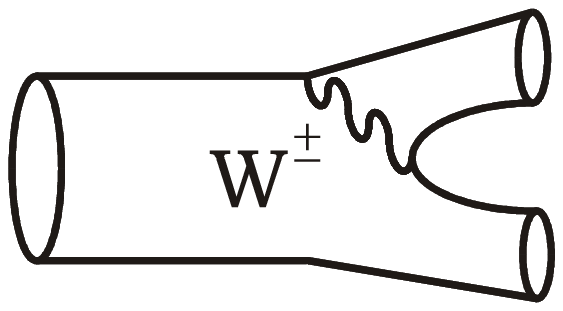}%
}%
&
{\includegraphics[
height=0.9115in,
width=1.0101in
]%
{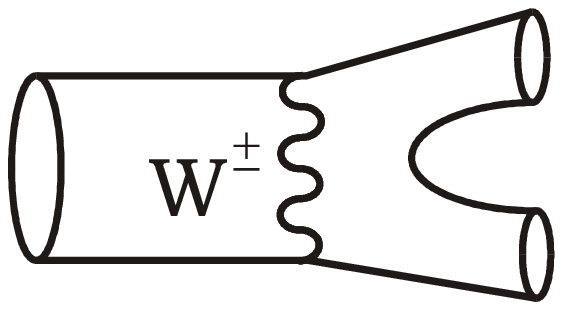}%
}%
\\
Tree $\left(  T\right)  $ & Color-suppressed $\left(  C\right)  $ & Exchange
$\left(  E\right)  $\\%
{\includegraphics[
height=0.9115in,
width=1.5004in
]%
{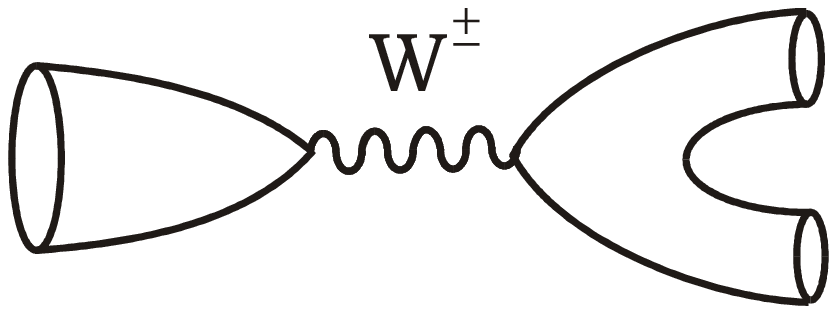}%
}%
&
{\includegraphics[
height=0.9115in,
width=1.0222in
]%
{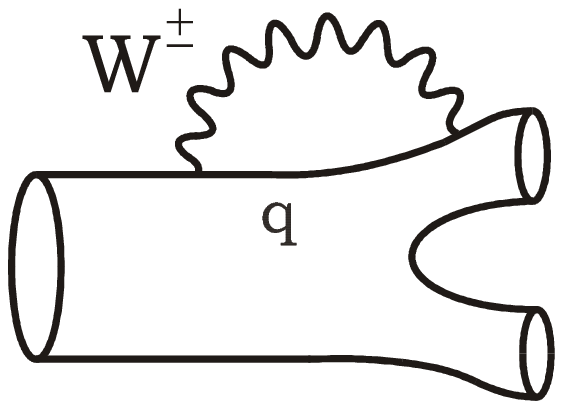}%
}%
&
{\includegraphics[
height=0.9115in,
width=1.3223in
]%
{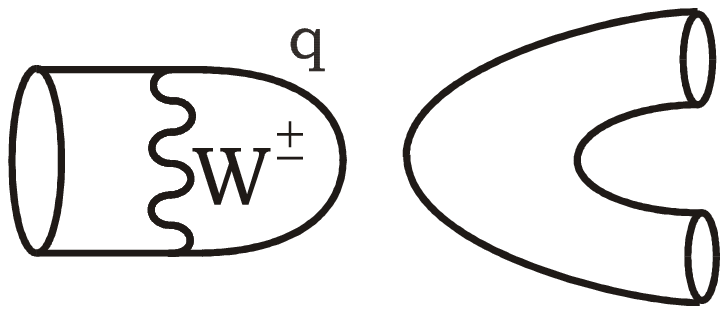}%
}%
\\
Annihilation $\left(  A\right)  $ & Penguin $\left(  P_{q}\right)  $ & Penguin
Annihilation $\left(  PA_{q}\right)  $%
\end{tabular}
\\
\\
\textbf{Figure 1:} Quark diagrams for non-singlet final states, $q=u,c,t$.
\end{tabular}
\]
In principle, second-order weak interaction topologies are very small, except
for those depicted in Fig.2 involving an enhancement factor $m_{top}^{2}%
/M_{Z}^{2}$.
\[%
\begin{tabular}
[c]{c}%
\begin{tabular}
[c]{cc}%
{\includegraphics[
height=0.9115in,
width=1.0966in
]%
{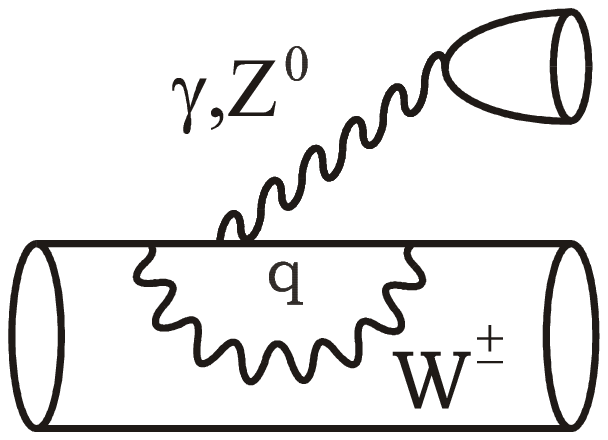}%
}%
&
{\includegraphics[
height=0.9115in,
width=1.017in
]%
{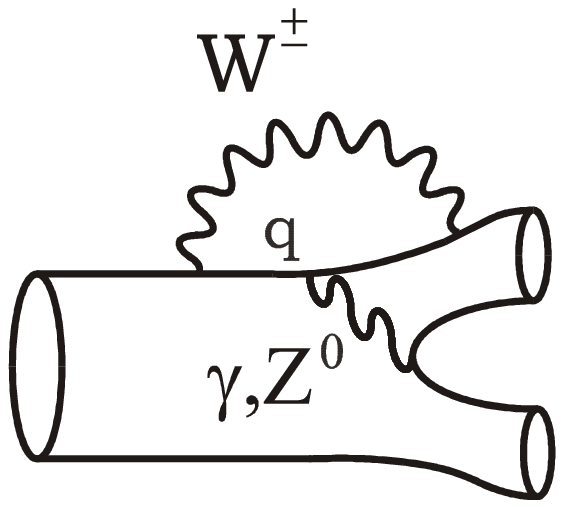}%
}%
\\
Electroweak Penguin $\left(  P_{q}^{EW}\right)  $ & $%
\begin{array}
[c]{c}%
\text{Color-suppressed}\\
\text{Electroweak Penguin}\left(  P_{C,q}^{EW}\right)
\end{array}
$%
\end{tabular}
\\
\\
\textbf{Figure 2:} Second-order weak interaction quark diagrams.
\end{tabular}
\]
The SU(3) symmetry is implemented at the level of bare decay amplitudes by
identifying quark diagrams that differ only by the interchange
$u\leftrightarrow d\leftrightarrow s$. The badly broken SU(4) symmetry is not
invoked and, consequently, quark diagrams for $B\rightarrow\left\{  D\bar
{D}\right\}  $ will be distinguished from $B\rightarrow\left\{  PP\right\}  $
ones by a subscript $D$.

Rescattering matrices have an interesting property of factorization. If each
of the strong phases is expressed as the sum of two parts%
\begin{equation}
\delta_{C_{i}}=\delta_{C_{i}}^{a}+\delta_{C_{i}}^{b}\;, \label{Eq5}%
\end{equation}
then, from the unitarity of $\sqrt{S}$ (see Eq.(\ref{Eq4}))
\begin{equation}
\sqrt{S}\left(  \delta_{C_{1}},...\right)  =\sqrt{S}(\delta_{C_{1}}%
^{a},...)\,.\,\sqrt{S}(\delta_{C_{1}}^{b},...) \label{Eq6}%
\end{equation}
Now, we can define \underline{effective} ''bare'' amplitudes containing the
$\delta_{C_{i}}^{b}$ part of the CP-conserving strong phases:%
\begin{equation}
W=\sqrt{S}\left(  \delta_{C_{1}},...\right)  \,.\,W_{b}\equiv\sqrt{S}%
(\delta_{C_{1}}^{a},...)\,.\,W_{b}^{eff} \label{Eq7}%
\end{equation}
In just the same way $W_{b}$ is parametrized in terms of real bare quark
diagrams $T,C,$..., the effective bare amplitudes $W_{b}^{eff}$ are decomposed
into complex effective quark diagram amplitudes $T^{eff},C^{eff},$...(i.e.
$W_{b}^{eff}=W_{b}\left(  T\rightarrow T^{eff},C\rightarrow C^{eff}%
,...\right)  $). It is important to realize that in the literature, quark
diagrams are often introduced at the level of $W_{b}^{eff}$. One should then
be very careful in understanding which part of the FSI they contain, so that
different approaches can be related. Finally, note that $CP(W_{b}^{eff}%
)\neq(W_{b}^{eff})^{\ast}$ since some CP-invariant strong phases are now
included into them (compare with Eqs.(1)).

\subsection{SU(3)-elastic Rescattering}

SU(3)-elasticity is enforced by taking definite SU(3) states as rescattering
eigenchannels. Taking as an example the set of decay amplitudes%
\begin{equation}
W=\left(
\begin{array}
[c]{l}%
B^{+}\rightarrow K^{+}\eta_{8}\\
B^{+}\rightarrow K^{+}\pi^{0}\\
B^{+}\rightarrow K^{0}\pi^{+}\\
B^{+}\rightarrow D_{s}^{+}\bar{D}^{0}%
\end{array}
\right)  \;, \label{Eq8}%
\end{equation}
the rescattering matrix $S$ is diagonal%
\begin{equation}
S_{diag}=diag\left(  e^{2i\delta_{27}},e^{2i\delta_{27}},e^{2i\delta_{8}%
},e^{2i\delta_{8}^{D}}\right)  \label{Eq9}%
\end{equation}
in the basis defined by%
\begin{equation}
\left(
\begin{array}
[c]{c}%
\left|  27,3/2,1/2,1\right\rangle \smallskip\\
\left|  27,1/2,1/2,1\right\rangle \smallskip\\
\left|  8,1/2,1/2,1\right\rangle \smallskip\\
\left|  8_{D},1/2,1/2,1\right\rangle
\end{array}
\right)  =\left(
\begin{array}
[c]{cccc}%
0 & -\sqrt{\frac{2}{3}} & \frac{1}{\sqrt{3}} & 0\\
-\frac{3}{\sqrt{10}} & -\frac{1}{\sqrt{30}} & -\frac{1}{\sqrt{15}} & 0\\
\frac{1}{\sqrt{10}} & -\sqrt{\frac{3}{10}} & -\sqrt{\frac{3}{5}} & 0\\
0 & 0 & 0 & 1
\end{array}
\right)  \left(
\begin{array}
[c]{l}%
\left\{  K^{+}\eta_{8}\right\}  \smallskip\\
\left\{  K^{+}\pi^{0}\right\}  \smallskip\\
\left\{  K^{0}\pi^{+}\right\}  \smallskip\\
\left\{  D_{s}^{+}\bar{D}^{0}\right\}
\end{array}
\right)  \label{Eq10}%
\end{equation}
Therefore, the SU(3)-elastic rescattering matrix is%
\begin{equation}
\sqrt{S_{SU\left(  3\right)  }}=\left(
\begin{array}
[c]{cccc}%
\frac{9e^{i\delta_{27}}+e^{i\delta_{8}}}{10} & \frac{\sqrt{3}\left(
e^{i\delta_{27}}-e^{i\delta_{8}}\right)  }{10} & \frac{\sqrt{3}\left(
e^{i\delta_{27}}-e^{i\delta_{8}}\right)  }{\sqrt{50}} & 0\\
\frac{\sqrt{3}\left(  e^{i\delta_{27}}-e^{i\delta_{8}}\right)  }{10} &
\frac{7e^{i\delta_{27}}+3e^{i\delta_{8}}}{10} & \frac{3\left(  e^{i\delta_{8}%
}-e^{i\delta_{27}}\right)  }{\sqrt{50}} & 0\\
\frac{\sqrt{3}\left(  e^{i\delta_{27}}-e^{i\delta_{8}}\right)  }{\sqrt{50}} &
\frac{3\left(  e^{i\delta_{8}}-e^{i\delta_{27}}\right)  }{\sqrt{50}} &
\frac{2e^{i\delta_{27}}+3e^{i\delta_{8}}}{5} & 0\\
0 & 0 & 0 & e^{i\delta_{8}^{D}}%
\end{array}
\right)  \label{Eq11}%
\end{equation}
See how SU(3)-elasticity fixes both the eigenphases Eq.(\ref{Eq9})
\textit{and} the eigenchannels Eq.(\ref{Eq10}): the $\left\{  D_{s}^{+}\bar
{D}^{0}\right\}  $ and $\left\{  PP\right\}  $ channels are not coupled and
$\sqrt{S_{SU\left(  3\right)  }}$ has a characteristic bloc-diagonal form.
Other sets of states are treated similarly (see \cite{MyPaper}).

Let us now concentrate on the $PP\rightleftharpoons PP$ part of $\sqrt
{S_{SU\left(  3\right)  }}$. If each strong phase is expressed as $\delta
_{i}=\delta_{i}^{l}+\delta_{i}^{s}$ with $i=1,8,27$ (one could think of the
FSI short and long-distance parts), we can define effective, complex quark
diagrams by absorbing the $\delta_{i}^{s}$ part of the rescattering
\begin{equation}
W_{b}^{eff}=\sqrt{S_{SU\left(  3\right)  }}\left(  \delta_{27}^{s},\delta
_{8}^{s},\delta_{1}^{s}\right)  W_{b} \label{Eq12}%
\end{equation}
Explicitly, effective quark diagrams can be expressed in terms of bare (real)
ones as (omitting electroweak penguins for now)%
\begin{equation}
\text{SU(3)-elasticity\ }\left\{
\begin{array}
[c]{l}%
X^{eff}=Xe^{i\delta_{8}^{s}}+\frac{e^{i\delta_{27}^{s}}-e^{i\delta_{8}^{s}}%
}{2}\left(  T+C\right)  \;\;\;\;\;\;\;\;\;\;X=T,C\\
Y^{eff}=Ye^{i\delta_{8}^{s}}-\frac{e^{i\delta_{27}^{s}}-e^{i\delta_{8}^{s}}%
}{10}\left(  T+C\right)  \;\;\;\;\;\;\;\;\;\;Y=E,A,P_{u}\\
P_{c,t}^{eff}=P_{c,t}e^{i\delta_{8}^{s}}\\
PA_{u}^{eff}=PA_{u}e^{i\delta_{1}^{s}}-\frac{e^{i\delta_{27}^{s}}%
-e^{i\delta_{8}^{s}}}{20}\left(  T+C\right)  +\frac{e^{i\delta_{1}^{s}%
}-e^{i\delta_{8}^{s}}}{12}\left(  3T-C+8E+8P_{u}\right) \\
PA_{c,t}^{eff}=PA_{c,t}e^{i\delta_{1}^{s}}+\frac{2(e^{i\delta_{1}^{s}%
}-e^{i\delta_{8}^{s}})}{3}P_{c,t}%
\end{array}
\right.  \label{Eq13}%
\end{equation}
The remaining $\delta_{i}^{l}$ part of the rescattering is accounted for by
acting with $\sqrt{S_{SU\left(  3\right)  }}\left(  \delta_{27}^{l},\delta
_{8}^{l},\delta_{1}^{l}\right)  $ on $W_{b}^{eff}$.

In the following, we shall not use Eqs.(\ref{Eq13}). What we want to point out
by writing them is that one should be careful when dealing with effective
quark diagrams since obviously, the rough scalings (penguins are discussed in
the next section)%
\begin{align}
1  &  :T\nonumber\\
\mathcal{O}\left(  \lambda\right)   &  :C\label{Eq14}\\
\mathcal{O}\left(  \lambda^{2}\right)   &  :E,A\nonumber
\end{align}
with $\lambda\approx1/5$ \cite{Gronau} may no longer be valid. For example,
for the $B^{+}\rightarrow K^{0}\pi^{+}$ mode, while $A$ is helicity-suppressed
and can be safely ignored, $A^{eff}$ receives unsuppressed contributions from
$T$ (provided $\delta_{27}^{s}-\delta_{8}^{s}$ is non-negligible), as already
emphasized in \cite{Annihilation}. Similarly, for $B^{0}\rightarrow\pi^{+}%
\pi^{-}$, $E$ and $PA$ can be neglected, but $E^{eff}$ and $PA^{eff}$ may not.
The Eqs.(\ref{Eq13}) illustrate in the specific SU(3)-elastic case a more
general fact: if one starts with a parametrization in terms of effective quark
diagrams without specifying anything about FSI, all the effective quark
diagrams are then arbitrary complex numbers and this costs much in terms of
free parameters (except, of course, if one has a definite computation scheme
for effective quark diagrams).

\subsection{Intermediate Charm Rescattering}

We now open the rescattering channel $D\bar{D}\rightleftharpoons PP$
\cite{OtherDD}. For the set of states Eq.(\ref{Eq8}), general rescattering
eigenchannels are defined by mixing $8$ with $8_{D}$ (for a discussion on the
formalism used here, see \cite{MyPaper}, \cite{PREVIOUS}):%
\begin{equation}
\left(
\begin{array}
[c]{c}%
C_{1}\left(  27,3/2\right) \\
C_{2}\left(  27,1/2\right) \\
C_{3}\left(  8,1/2\right) \\
C_{4}\left(  8,1/2\right)
\end{array}
\right)  =\left(
\begin{array}
[c]{cccc}%
1 & 0 & 0 & 0\\
0 & 1 & 0 & 0\\
0 & 0 & \cos\chi_{8} & \sin\chi_{8}\\
0 & 0 & -\sin\chi_{8} & \cos\chi_{8}%
\end{array}
\right)  \left(
\begin{array}
[c]{c}%
\left|  27,3/2,1/2,1\right\rangle \\
\left|  27,1/2,1/2,1\right\rangle \\
\left|  8,1/2,1/2,1\right\rangle \\
\left|  8_{D},1/2,1/2,1\right\rangle
\end{array}
\right)  \label{Eq15}%
\end{equation}
The rescattering matrix in the $C_{i}$ basis is%
\begin{equation}
S_{diag}=diag\left(  e^{2i\delta_{C_{1}}},e^{2i\delta_{C_{2}}},e^{2i\delta
_{C_{3}}},e^{2i\delta_{C_{4}}}\right)  \approx diag\left(  e^{2i\delta_{27}%
},e^{2i\delta_{27}},e^{2i\delta_{8}},e^{2i\delta_{8}^{D}}\right)  \label{Eq16}%
\end{equation}
since the mixing angle $\chi_{8}$ should be small (see Eq.(\ref{Eq4b})). In
the physical state basis, we then find
\begin{align}
\sqrt{S_{\chi_{8}}}  &  =\sqrt{S_{SU\left(  3\right)  }}+\sin\left(  2\chi
_{8}\right)  \frac{e^{i\delta_{8}}-e^{i\delta_{8}^{D}}}{2\sqrt{10}}\left(
\begin{tabular}
[c]{llll}%
$0$ & $0$ & $0$ & $1$\\
$0$ & $0$ & $0$ & $-\sqrt{3}$\\
$0$ & $0$ & $0$ & $-\sqrt{6}$\\
$1$ & $-\sqrt{3}$ & $-\sqrt{6}$ & $0$%
\end{tabular}
\right) \nonumber\\
&  +\sin^{2}\left(  \chi_{8}\right)  \frac{e^{i\delta_{8}}-e^{i\delta_{8}^{D}%
}}{10}\left(
\begin{tabular}
[c]{llll}%
$-1$ & $\sqrt{3}$ & $\sqrt{6}$ & $0$\\
$\sqrt{3}$ & $-3$ & $-3\sqrt{2}$ & $0$\\
$\sqrt{6}$ & $-3\sqrt{2}$ & $-6$ & $0$\\
$0$ & $0$ & $0$ & $10$%
\end{tabular}
\right)  \label{Eq17}%
\end{align}
with $\sqrt{S_{SU\left(  3\right)  }}$ given in Eq.(\ref{Eq11}). In physical
terms, one can understand $\chi_{8}$ as a $D\bar{D}$-$PP$ ''coupling
constant'': the $\sin\left(  2\chi_{8}\right)  $ term describes the $D\bar{D}$
pollution of $PP$ states (which proceeds through $D\bar{D}\overset{\chi_{8}%
}{\rightarrow}PP,D\bar{D}\overset{\chi_{8}}{\rightarrow}PP\overset{\chi_{8}%
}{\rightarrow}D\bar{D}\overset{\chi_{8}}{\rightarrow}PP,$ etc) while the
$\sin^{2}\left(  \chi_{8}\right)  $ terms account for ''inelastic'' distortion
of the rescattering among $\left\{  PP\right\}  $ states due to the exchange
of probability with the $D\bar{D}$ channel ($PP\overset{\chi_{8}}{\rightarrow
}D\bar{D}\overset{\chi_{8}}{\rightarrow}PP,$ etc). The common factor
$e^{i\delta_{8}}-e^{i\delta_{8}^{D}}$ acts as a kinematical suppression since
it decreases as the relative momentum between decay products increases
($\delta_{8}=\delta_{8}^{D}=0$ in the limit $M_{B}\rightarrow\infty$
\cite{MBInfinite}). Finally, note well that SU(3) remains exact, and that the
rescattering is elastic (conserved total probability) for the full set of
states $\left\{  PP,D\bar{D}\right\}  $ since $\sqrt{S_{\chi_{8}}}$ is unitary
for all $\chi_{8}$.

Other set of states are treated similarly (the mixings among $Y=0,T_{3}=0$
states is described in Appendix A). All in all, six parameters are needed to
describe all the rescattering mixings (one of the strong phase can be
eliminated)%
\begin{equation}%
\begin{tabular}
[c]{rcclcll}%
$PP:\left(  8\otimes8\right)  _{S}=$ & $27$ & $\oplus$ & $8$ & $\oplus$ & $1 $
& $\;\;\rightarrow\delta_{27},\delta_{8},\delta_{1}$\\
&  &  & $\updownarrow\chi_{8}$ &  & $\updownarrow\chi_{1}$ & \\
$D\bar{D}:\;\;3\otimes\bar{3}\;\;\;=$ &  &  & $8_{D}$ & $\oplus$ & $1_{D}$ &
$\;\;\rightarrow\delta_{8}^{D},\delta_{1}^{D}$%
\end{tabular}
\label{Eq18}%
\end{equation}
We expect that $\chi_{8},\chi_{1}<<1$ since $D\bar{D}\rightarrow PP$ proceeds
through the annihilation of the $c\bar{c}$ pair into light quarks, and
$\delta_{27},\delta_{8},\delta_{1}<\delta_{8}^{D},\delta_{1}^{D}$ since there
is less phase-space for $D\bar{D}$ than for $PP$. As said before,
SU(4)-elasticity is not implemented: it would fix the $\chi_{i}$ to some large
values and introduce relations between $\delta_{27},\delta_{8},\delta_{1}$ and
$\delta_{8}^{D},\delta_{1}^{D}$, leading to $Br\left(  B\rightarrow PP\right)
\approx Br\left(  B\rightarrow D\bar{D}\right)  $ (see Eq.(\ref{Eq4b})). Let
us stress again that SU(4) is used at no stage of the present analysis.

The factorization property of $\sqrt{S}$ can now be used to separate the
$D\bar{D}\rightleftharpoons PP$ rescattering effects from the SU(3)-elastic
part. For the set of states Eq.(\ref{Eq8}), this is achieved by splitting the
phases as (see Eq.(\ref{Eq17}))%
\begin{equation}
\sqrt{S_{\chi_{8}}}\left(  \delta_{27},\delta_{8},\delta_{8}^{D}\right)
=\underset{%
\begin{array}
[c]{c}%
PP\rightleftharpoons PP\\
D\bar{D}\rightleftharpoons D\bar{D}%
\end{array}
}{\underbrace{\sqrt{S_{\chi_{8}}}\left(  \delta_{27},\delta_{8},\delta
_{8}\right)  }}\;.\;\underset{%
\begin{array}
[c]{c}%
D\bar{D}\rightleftharpoons PP
\end{array}
}{\underbrace{\sqrt{S_{\chi_{8}}}\left(  0,0,\delta_{8}^{D}-\delta_{8}\right)
}} \label{Eq19}%
\end{equation}
The first factor does not depend on $\chi_{8}$ and is related to the
SU(3)-elastic rescattering matrix Eq.(\ref{Eq11}) as $\sqrt{S_{\chi_{8}}%
}\left(  \delta_{27},\delta_{8},\delta_{8}\right)  =\sqrt{S_{SU\left(
3\right)  }}\left(  \delta_{27},\delta_{8},\delta_{8}\right)  $, while the
second factor contains all the effects of $D\bar{D}\rightleftharpoons PP$
mixing and becomes trivial if either $\delta_{8}^{D}=\delta_{8}$ or $\chi
_{8}=0$. Proceeding similarly with the other multiplets, we now absorb the
$D\bar{D}\rightleftharpoons PP$ part of the rescattering into effective quark
diagrams%
\begin{equation}
\text{On-shell }c\bar{c}\text{ FSI\ }\left\{
\begin{array}
[c]{l}%
P_{c}^{eff}=P_{c}-\beta\\
PA_{c}^{eff}=PA_{c}+\frac{2}{3}\beta+\alpha
\end{array}
\right.  \label{Eq20}%
\end{equation}
with
\begin{equation}
\alpha=\frac{1}{2\sqrt{3}}\left(  1-e^{i(\delta_{1}^{D}-\delta_{1})}\right)
\chi_{1}T_{D},\;\;\;\beta=\sqrt{\frac{3}{5}}\left(  1-e^{i(\delta_{8}%
^{D}-\delta_{8})}\right)  \chi_{8}T_{D} \label{Eq21}%
\end{equation}
To reach these expressions, we have retained only the dominant $T_{D}$
contribution to $B\rightarrow D\overline{D}$ decay amplitudes (see Fig.1) and
the first order in $\chi_{i}$ (so that $PP\rightarrow D\overline{D}%
\rightarrow...\rightarrow PP$ effects are neglected, see Eq.(\ref{Eq17})).
Importantly, all the other quark diagrams are unaffected by $D\bar
{D}\rightleftharpoons PP$. Since the $P_{c}$ and $PA_{c}$ quark diagrams are
precisely the ones involving the charm quark, we interpret Eq.(\ref{Eq20}) as
a hadronic representation for on-shell intermediate $c\bar{c}$ quarks, as
previously mentioned in \cite{PREVIOUS} (see Fig.3).%
\begin{align*}
&
{\includegraphics[
height=0.9444in,
width=1.5748in
]%
{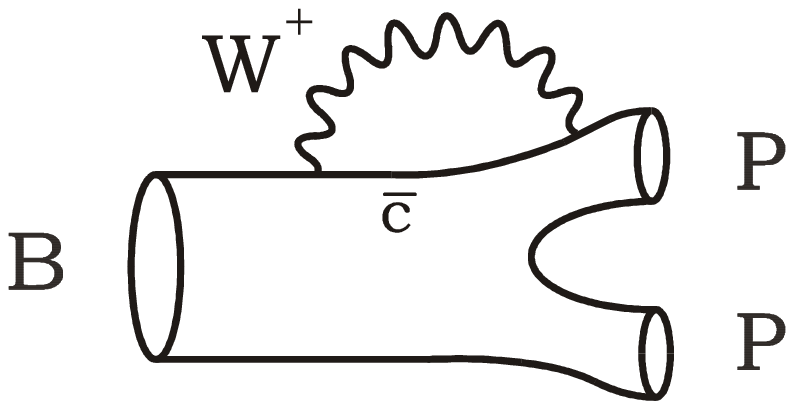}%
}%
\;\;\;\;\;\;\;\;\;\;\;\;%
{\includegraphics[
height=0.9409in,
width=2.1664in
]%
{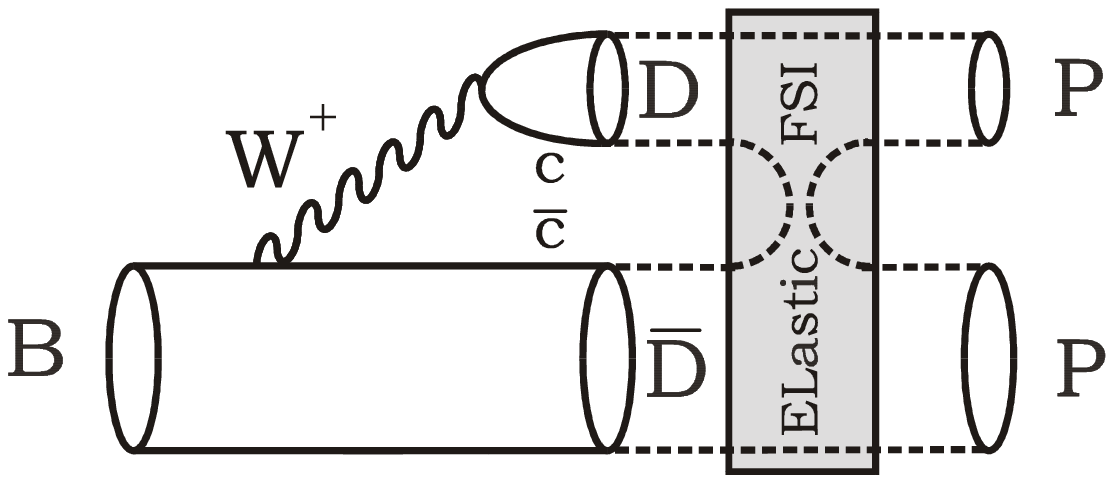}%
}%
\\
&  \;\;%
\begin{array}
[c]{l}%
\text{\textbf{Figure 3: }The standard quark-level penguin amplitude }%
P_{c}\text{ and }\\
\text{the hadronic-level penguin-like }B\rightarrow\left\{  D\bar{D}\right\}
\rightarrow PP\text{ contribution.}%
\end{array}
\end{align*}

In general, because of renormalization of the penguin loop, it is desirable to
use the unitarity of the CKM matrix to write ($\lambda_{q}\equiv V_{qb}^{\ast
}V_{qd\left(  s\right)  }$)%
\begin{equation}
\lambda_{u}P_{u}+\lambda_{c}P_{c}+\lambda_{t}P_{t}=\lambda_{u}\left(
P_{u}-P_{c}\right)  +\lambda_{t}\left(  P_{t}-P_{c}\right)  \equiv\lambda
_{u}\left(  P_{u-c}\right)  +\lambda_{t}\left(  P_{t-c}\right)  \label{Eq22}%
\end{equation}
At the $B$-mass scale, we expect that $P_{u-c}$ is small since $m_{u}%
,m_{c}<<m_{b}$. Interestingly, when on-shell intermediate $c\bar{c}$ is
treated at the hadronic level, the $P_{c}$ short-distance absorptive part has
to be discarded and to a large extent the cancellation between $P_{u}$ and
$P_{c}$ is preserved. Therefore, one can expect the rough scaling of the
dominant penguin contributions \cite{Gronau}
\begin{align}
1  &  :T\nonumber\\
\mathcal{O}\left(  \lambda\right)   &  :P_{t-c}\label{Eq23}\\
\mathcal{O}\left(  \lambda^{2}\right)   &  :P_{t-c}^{EW}\nonumber\\
\mathcal{O}\left(  \lambda^{3}\right)   &  :P_{C,t-c}^{EW},PA_{t-c}\nonumber
\end{align}
and $\left|  X_{u-c}\right|  <\left|  X_{t-c}\right|  $ with $X=P,PA,P^{EW}%
,P_{C}^{EW}$ to be valid. In the following, $X_{u-c}$ will be neglected and
the effects of intermediate charm will be accounted for by defining%
\begin{equation}
\text{On-shell }c\bar{c}\text{ FSI\ }\left\{
\begin{array}
[c]{l}%
\lambda_{t}P_{t-c}^{eff}=\lambda_{t}P_{t-c}-\lambda_{c}\beta\\
\lambda_{t}PA_{t-c}^{eff}=\lambda_{t}PA_{t-c}+\lambda_{c}\left(  \frac{2}%
{3}\beta+\alpha\right)
\end{array}
\right.  \label{Eq24}%
\end{equation}
Note well that $P_{t-c}^{eff}$ and $PA_{t-c}^{eff}$ involve a combination of
CP-conserving and CP-violating phases.

Our method for implementing intermediate on-shell charm has semi-inclusive
features. To see this, note first that all the developments above can be
repeated replacing $D\bar{D}$ by $D^{\ast}\bar{D}^{\ast},D^{\ast\ast}\bar
{D}^{\ast\ast},...$ Now, taking into account the effects on $B\rightarrow PP$
of rescatterings from all the charmed meson states amounts to the definitions
Eqs.(\ref{Eq20}) or Eqs.(\ref{Eq24}) with $\alpha,\beta$ replaced by%
\begin{equation}
\widetilde{\alpha}=\frac{1}{2\sqrt{3}}F_{1},\;\;\;\widetilde{\beta}%
=\sqrt{\frac{3}{5}}F_{8}\text{,\ \ \ \ \ }F_{k}=\sum_{i=D,D^{\ast},...}\left(
1-e^{i(\delta_{k}^{i}-\delta_{k})}\right)  \chi_{k}^{i}T_{i} \label{Eq25}%
\end{equation}
and this holds even if $D\bar{D}\rightleftharpoons D^{\ast}\bar{D}^{\ast},...$
rescatterings are present thanks to the factorization property Eq.(\ref{Eq6})
or Eq.(\ref{Eq19}). Finally, since in any case, the complex numbers
$\widetilde{\alpha}$ and $\widetilde{\beta}$ can be parametrized as in
Eq.(\ref{Eq21}), this change is irrelevant (except, of course, that $\chi$ and
$T_{D}$ can no longer be determined separately). In other words, as long as
the $\chi_{i}$ are small (i.e. when the last term of Eq.(\ref{Eq17}) can be
dropped), Watson's theorem implies that no matter the precise physics in the
charmed sector, its impact on $B\rightarrow PP$ will always amount to a
redefinition of the penguin amplitudes $P_{c}$ and $PA_{c}$ (or $P_{t-c}$ and
$PA_{t-c}$). In the following, for simplicity, we assume that $D\bar
{D}\rightleftharpoons PP$ dominates so as to give estimations of $\chi_{i}$.
Finally, it is also straightforward to apply the formalism to charmonium modes
$P\eta_{c}$ and/or $\eta_{0}\eta_{c}$ (see \cite{Barshay}), simply by
substituting $T_{D}$ by the $C_{\eta_{c}}$ quark diagram contributing to
$P\eta_{c}$ in $\beta$, and $T_{D}$ by the $C_{\eta^{c}}^{\prime}$
contributing to $\eta_{0}\eta_{c}$ in $\alpha$.

Let us conclude this section by a more technical comment. If we note $\bar
{3}_{c}$ the $\bar{b}c.\bar{c}d$\ or $\bar{b}c.\bar{c}s$\ part of the weak
Hamiltonian (see \cite{MyPaper}), $P_{c}$, $PA_{c}$ and the definitions
Eq.(\ref{Eq20}) are expressed in terms of SU(3) reduced matrix elements as%
\begin{equation}
\left\{
\begin{array}
[c]{l}%
P_{c}=-\left\langle 8\left|  \bar{3}_{c}\right|  3\right\rangle \\
PA_{c}=\frac{2}{3}\left\langle 8\left|  \bar{3}_{c}\right|  3\right\rangle
+\left\langle 1\left|  \bar{3}_{c}\right|  3\right\rangle
\end{array}
\right.  \;\;\rightarrow\;\;\left\{
\begin{array}
[c]{c}%
\left\langle 8\left|  \bar{3}_{c}\right|  3\right\rangle ^{eff}=\left\langle
8\left|  \bar{3}_{c}\right|  3\right\rangle +\beta\\
\left\langle 1\left|  \bar{3}_{c}\right|  3\right\rangle ^{eff}=\left\langle
1\left|  \bar{3}_{c}\right|  3\right\rangle +\alpha
\end{array}
\right.  \label{Eq25b}%
\end{equation}
As expected, the $\alpha\left(  \beta\right)  $ term involving $\chi
_{1}\left(  \chi_{8}\right)  $ contributes only to the reduced matrix element
involving the $1\left(  8\right)  $ final state, respectively.

\subsection{Final Parametrization}

All the pieces can now be put together to construct the final parametrizations
of $B\rightarrow PP$ decay amplitudes assuming enlarged SU(3)-elasticity. From
the previous sections:

\begin{quote}
1- All $PP\rightleftharpoons PP$ rescattering effects should be contained in
$\sqrt{S}$, to preserve the scalings between quark diagram amplitudes, and
will be treated assuming exact SU(3) (see Eqs.(\ref{Eq13}) and Eqs.(\ref{Eq14})).

2- Intermediate charm can be treated at the hadronic level as $D\bar
{D}\rightleftharpoons PP$ rescattering effects, and absorbed into $P_{t-c}$
and $PA_{t-c}$ (see Eq.(\ref{Eq24})).
\end{quote}

The first point is important since it allows us to reduce the number of free
parameters. Indeed, combining Eqs.(\ref{Eq14}) and Eqs.(\ref{Eq23}) with CKM
coefficients, it appears as sufficient to consider only four topologies
($\lambda_{q}^{q^{\prime}}\equiv V_{qb}^{\ast}V_{qq^{\prime}}$)%
\begin{align}
\Delta S  &  =0:%
\begin{array}
[c]{cc}%
\lambda_{u}^{d}\,T,\;\;\;\;\; & \lambda_{t}^{d}\,P_{t-c},\lambda_{u}^{d}\,C
\end{array}
\label{Eq26}\\
\Delta S  &  =1:%
\begin{array}
[c]{cc}%
\lambda_{t}^{s}P_{t-c},\;\;\; & \lambda_{u}^{s}\,T,\lambda_{t}^{s}%
\,P_{t-c}^{EW}%
\end{array}
\nonumber
\end{align}
since in any case, SU(3) breaking effects in the dominant $T$ and $P_{t-c}$
contributions should be greater than $\mathcal{O}\left(  \lambda^{2}\right)
\sim4\%$.

Once $PA$ is neglected, it seems justified to set also (see Eq.(\ref{Eq24}))%
\begin{equation}
\frac{2}{3}\beta+\alpha=0\Leftrightarrow\left\{
\begin{array}
[c]{l}%
\delta_{1}^{D}-\delta_{1}=\delta_{8}^{D}-\delta_{8}\\
\chi\equiv\chi_{8}=-\frac{\sqrt{5}}{4}\chi_{1}%
\end{array}
\right.  \label{Eq27}%
\end{equation}
since this combination describes rescatterings that proceeds through the
vacuum (Eq.(\ref{Eq27}) may seem to imply a fine tuning between $\alpha$ and
$\beta$, but this is equivalent to the usual assumption that $PA_{c}$ is
small, see Eqs.(\ref{Eq25b})).

There remain eight free parameters
\begin{align}
\text{QD amplitudes}\text{: \ }  &  T,C,P_{t-c},P_{t-c}^{EW}\nonumber\\
\text{SU(3) rescatterings:\ }  &  \delta_{27}-\delta_{1},\delta_{8}-\delta
_{1}\label{Eq28}\\
\text{On-shell }c\bar{c}\text{ FSI}\text{:\ }  &  \chi T_{D},\delta_{8}%
^{D}-\delta_{1}\nonumber
\end{align}
The expressions for decay amplitudes are collected in Appendix B, in which the
definition Eq.(\ref{Eq24}) is used, $\delta_{1}$ is set to zero and CKM
coefficients are omitted. Also, the color-allowed electroweak penguin has to
be introduced as $C\rightarrow C+P_{t-c}^{EW}$ \cite{Gronau}. Decay amplitudes
for $B\rightarrow D\bar{D}$, assuming that $D\bar{D},PP\rightleftharpoons
D^{\ast}\bar{D}^{\ast},...$ are negligible, are also given. In that case,
measurements of $B\rightarrow D\bar{D}$ branchings can serve to determine both
$T_{D}$ and the strong phase $\delta_{8}-\delta_{1}$ of $PP\rightleftharpoons
PP$ rescattering. This fact originates in the coherence requirement
Eq.(\ref{Eq27}), which implies (provided $\chi\neq0$)%
\begin{equation}
\frac{Br\left(  B^{0}\rightarrow D^{+}D^{-}\right)  \Gamma\left(
B^{0}\right)  }{Br\left(  B^{+}\rightarrow D^{+}\bar{D}^{0}\right)
\Gamma\left(  B^{+}\right)  }=\frac{5+4\cos\left(  \delta_{8}^{D}-\delta
_{1}^{D}\right)  }{9}=\frac{5+4\cos\left(  \delta_{8}-\delta_{1}\right)  }{9}
\label{Eq29}%
\end{equation}

\section{Numerical Examples}

For our numerical examples, we consider only one source of rescattering:
either SU(3)-elastic ones ($PP\rightleftharpoons PP$) or on-shell intermediate
charm ($D\bar{D}\rightleftharpoons PP$). The sets of free parameters in each
case are
\begin{equation}%
\begin{tabular}
[c]{cccc}%
$\text{Amplitudes:\ \ }$ & \multicolumn{3}{c}{$T,P_{t-c},C,P_{t-c}^{EW}$}\\
& $\swarrow$ &  & $\searrow$\\
$\text{Rescattering:\ \ }$ & $PP\rightleftharpoons PP$ &  & $D\bar
{D}\rightleftharpoons PP$\\
& $\delta_{27}-\delta_{1},\delta_{8}-\delta_{1}$ &  & $\chi T_{D},\delta
_{8}^{D}-\delta_{1}$%
\end{tabular}
\label{Ex1}%
\end{equation}
with, in addition, the weak angle $\gamma$. To leading order, one can further
neglect $C$ and $P_{t-c}^{EW}$. Though the formalism allows for general
analyses, the precision of the experimental data is not yet sufficient for
meaningful combined analyses.

For each fit, we give the values of the parameters in the text, while the
corresponding values for branchings (Table V), $A_{CP}$ (Table VI) and
$S_{f\bar{f}}$ (Table VII) are in Appendix C, along with the details of our
fitting procedure.

\subsection{SU(3)-elasticity (no $D\bar{D}\rightleftharpoons PP$)}

For the first example, we take the $T$ and $P_{t-c}$ amplitudes and
$\delta_{8}-\delta_{1}$, $\delta_{27}-\delta_{1}$ strong phases (so we set
$\chi=0$). For various (fixed) values of the weak angle $\gamma$ we find%
\[%
\begin{tabular}
[c]{c}%
\begin{tabular}
[c]{|lccc|}\hline
$\gamma$ & $60%
{{}^\circ}%
$ & $80%
{{}^\circ}%
$ & $100%
{{}^\circ}%
$\\\hline\hline
$\chi_{\min}^{2}$ & $18.3$ & $16.3$ & $17.7$\\\hline
$T$ & $0.65$ & $0.70$ & $0.76$\\
$P_{t-c}$ & $0.11$ & $0.11$ & $0.10$\\\hline
$\delta_{8}-\delta_{1}$ & $47%
{{}^\circ}%
$ & $52%
{{}^\circ}%
$ & $62%
{{}^\circ}%
$\\
$\delta_{27}-\delta_{1}$ & $100%
{{}^\circ}%
$ & $98%
{{}^\circ}%
$ & $101%
{{}^\circ}%
$\\\hline
\end{tabular}
\\
$\text{\textbf{Table\thinspace I:} }T,P_{t-c}$, SU(3)-elasticity.
\end{tabular}
\]
Quark diagram amplitudes given in the table are dimensionless and produce
$B^{0}$ branchings.

Under this SU(3)-elastic parametrization, the $B\rightarrow\pi K$ branchings
are dominated by the $P_{t-c}$ penguin amplitude and verify%
\begin{equation}
Br\left(  \pi^{0}K^{+}\right)  :Br\left(  \pi^{+}K^{0}\right)  :Br\left(
\pi^{0}K^{0}\right)  :Br\left(  \pi^{-}K^{+}\right)  \approx1/2:1:1/2:1
\label{Ex2}%
\end{equation}
While the pattern of direct CP-asymmetries is (see \cite{PREVIOUS})%
\begin{equation}
A_{\pi^{0}K^{+}}:A_{\pi^{+}K^{0}}:A_{\pi^{0}K^{0}}:A_{\pi^{-}K^{+}}%
\approx2:-\frac{1}{2}:-\frac{3}{2}:1 \label{Ex3}%
\end{equation}
no matter the strong phases.

As already found in \cite{Hou}, if the current CP-asymmetry measurement
central values are to be trusted, they require large $PP\rightleftharpoons PP$
phases. Then, in the $B\rightarrow\pi\pi$ sector, dominated by $T$, large
rescatterings occur and the $\pi^{0}\pi^{0}$ mode is fed from $B\rightarrow
\left\{  \pi^{+}\pi^{-}\right\}  $. The resulting pattern is close to
\begin{equation}
Br\left(  \pi^{+}\pi^{0}\right)  :Br\left(  \pi^{+}\pi^{-}\right)  :Br\left(
\pi^{0}\pi^{0}\right)  \approx1:1:1/2 \label{Ex4}%
\end{equation}
Typically, $Br\left(  B^{0}\rightarrow\pi^{0}\pi^{0}\right)  >1.5\times
10^{-6}$ under SU(3)-elasticity. For $\pi^{+}\pi^{-}$ time-dependent
asymmetry, both $A_{\pi^{+}\pi^{-}}$ and $S_{\pi^{+}\pi^{-}}$ are found
compatible with current experimental data (see appendix C):%
\[%
\begin{tabular}
[c]{c}%
{\includegraphics[
height=1.075in,
width=5.7009in
]%
{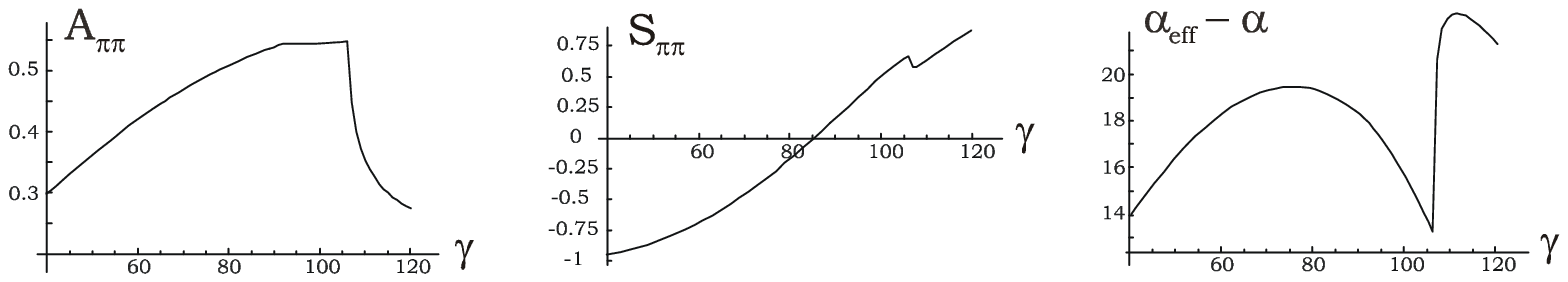}%
}%
\\
\textbf{Figure 4:} Fit result for $A_{\pi^{+}\pi^{-}},S_{\pi^{+}\pi^{-}}$ and
$\alpha_{eff}$ under SU(3)-elasticity.
\end{tabular}
\]
$A_{\pi^{+}\pi^{-}}$ is not very sensitive to $\gamma$ and stays in the range
0.3 to 0.5. On the other hand, $S_{\pi^{+}\pi^{-}}$ varies much, and it
appears that the present Belle measurement exclude $\gamma>90%
{{}^\circ}%
$ to $3\sigma$. However, the dependence of $S_{\pi^{+}\pi^{-}}$ on $\gamma$ is
rather trivial. Defining the parameter $\alpha_{eff}$ as%
\begin{equation}
S_{\pi^{+}\pi^{-}}=\sqrt{1-A_{\pi^{+}\pi^{-}}^{2}}\sin\left(  2\alpha
_{eff}\right)  \label{Ex5}%
\end{equation}
such that $\alpha_{eff}=\alpha\equiv\pi-\beta-\gamma$ if only $T$ contributes
to $B^{0}\rightarrow\pi^{+}\pi^{-}$, the last graph shows that the deviation
$\alpha_{eff}-\alpha$ stays small compared to $\gamma$.

Finally, for $B\rightarrow K\bar{K}$, no clear pattern emerges. However, one
can note that because of large rescatterings from $B\rightarrow\left\{
\pi^{+}\pi^{-}\right\}  $, the $B^{+}\rightarrow K^{+}K^{-}$ branching is
saturating its present experimental upper bound. Also, the $B^{+}\rightarrow
K^{+}\bar{K}^{0}$ and $B^{0}\rightarrow K^{0}\bar{K}^{0}$ direct
CP-asymmetries tend to be sizeable, between $-1$ and $-0.5$.

Just for illustration, if we now introduce the subleading $C$ and
$P_{t-c}^{EW}$ topologies, the best-fit values are
\[%
\begin{tabular}
[c]{c}%
\begin{tabular}
[c]{|lccc|}\hline
$\gamma$ & $60%
{{}^\circ}%
$ & $80%
{{}^\circ}%
$ & $100%
{{}^\circ}%
$\\\hline\hline
$\chi_{\min}^{2}$ & $10.1$ & $10.5$ & $15.4$\\\hline
$T$ & $0.60$ & $0.66$ & $0.67$\\
$C$ & $0.21$ & $0.17$ & $0.17$\\
$P_{t-c}$ & $0.11$ & $0.11$ & $0.10$\\
$P_{t-c}^{EW}$ & $0.011$ & $0.008$ & $0.005$\\\hline
$\delta_{8}-\delta_{1}$ & $53%
{{}^\circ}%
$ & $57%
{{}^\circ}%
$ & $32%
{{}^\circ}%
$\\
$\delta_{27}-\delta_{1}$ & $88%
{{}^\circ}%
$ & $89%
{{}^\circ}%
$ & $59%
{{}^\circ}%
$\\\hline
\end{tabular}
\\
$\text{\textbf{Table\thinspace II:} }T,P_{t-c},C,P_{t-c}^{EW}$,
SU(3)-elasticity.
\end{tabular}
\]
\newline The most noticeable feature is a reduction of the strong phases, at
the cost of a quite large $C$ amplitude, and a significant decrease in
$\chi_{\min}^{2}$.

To within $20\%$, the introduction of the additional quark diagram amplitudes
does not modify the patterns described above. For example, the ratios of
CP-asymmetries in the $\pi K$ sector are now%
\begin{equation}
A_{\pi^{0}K^{+}}:A_{\pi^{+}K^{0}}:A_{\pi^{0}K^{0}}:A_{\pi^{-}K^{+}}%
=2+\frac{8}{5}\kappa_{-1}:-\frac{1}{2}+\frac{3}{5}\kappa_{2/3}:-\frac{3}%
{2}+3\kappa_{0}:1 \label{Ex6}%
\end{equation}
with, to leading order in $\left|  \lambda_{u}^{s}\right|  T/\left|
\lambda_{t}^{s}\right|  P_{t-c}$ and $P_{t-c}^{EW}/P_{t-c}$
\begin{align}
\kappa_{\alpha}  &  =\cos\gamma\frac{\left|  \lambda_{u}^{s}\right|
T}{\left|  \lambda_{t}^{s}\right|  P_{t-c}}\left(  \cos\left(  \delta
_{8}-\delta_{27}\right)  \left(  1+\frac{C}{T}\right)  +\alpha\left(
1-\frac{3C}{2T}\right)  \right) \label{Ex7}\\
&  -\frac{P_{t-c}^{EW}}{P_{t-c}}\left(  \cos\left(  \delta_{8}-\delta
_{27}\right)  -\frac{\alpha}{4}\left(  1+\frac{5C}{C+T}\right)  \right)
\nonumber
\end{align}
For the range of parameter values in Table II, these corrections are of at
most $15\%$ (see Table VI).

To close this section, it should be noted that the SU(3)-elastic constraint
$A_{\pi^{0}K^{+}}=2A_{\pi^{-}K^{+}}$, though consistent, is not favored by the
present measurements of $A_{\pi^{-}K^{+}}$ by Belle, BaBar and $A_{\pi
^{0}K^{+}}$ by Belle:%
\[%
\begin{tabular}
[c]{ccc}
& Belle & BaBar\\\hline
$A_{\pi^{-}K^{+}}$ & $-0.088\pm0.035\pm0.018$\cite{Fry} & $-0.107\pm
0.041\pm0.013$\cite{Fry}\\
$A_{\pi^{0}K^{+}}$ & $0.23\pm0.11_{-0.04}^{+0.01}$\cite{Belle} &
$-0.09\pm0.09\pm0.01$\cite{BaBar}%
\end{tabular}
\]
Further, as shown in \cite{PREVIOUS}, this constraint is the same under
SU(2)-elasticity. More precise measurements are necessary to draw any
conclusion on the relevance of SU(N)-elastic rescatterings for $B\rightarrow
PP$.

\subsection{On-shell c\={c} rescattering (no $PP\rightleftharpoons PP$)}

For our second example, we take again the $T$ and $P_{t-c}$ amplitudes, but
$\delta_{8}^{D}-\delta_{8}$ as the only strong phase. In addition, we have the
$T_{D}$ amplitude and the mixing angle $\chi$. As a first step, since it is
always the combination $\left(  \chi T_{D}\right)  $ which appears in
$B\rightarrow PP$ decays, we can use the measurements \cite{PDG2002}
$Br\left(  B^{0}\rightarrow D^{-}D_{s}^{+}\right)  =\left(  0.8\pm0.3\right)
\%$ and $Br\left(  B^{+}\rightarrow\bar{D}^{0}D_{s}^{+}\right)  =\left(
1.3\pm0.4\right)  \%$ to fix $T_{D}=2.43$ which gives%
\begin{equation}
Br^{th}\left(  B^{0}\rightarrow D^{-}D_{s}^{+}\right)  =Br^{th}\left(
B^{+}\rightarrow\bar{D}^{0}D_{s}^{+}\right)  =0.98\% \label{Ex8}%
\end{equation}
For various weak angle $\gamma$ we find%
\[%
\begin{tabular}
[c]{c}%
\begin{tabular}
[c]{|lccc|}\hline
$\gamma$ & $60%
{{}^\circ}%
$ & $80%
{{}^\circ}%
$ & $100%
{{}^\circ}%
$\\\hline\hline
$\chi_{\min}^{2}$ & $40$ & $32$ & $26$\\\hline
$T$ & $0.50$ & $0.57$ & $0.66$\\
$P_{t-c}$ & $0.09$ & $0.10$ & $0.10$\\\hline
$\chi$ & $3.3%
{{}^\circ}%
$ & $2.7%
{{}^\circ}%
$ & $2.6%
{{}^\circ}%
$\\\hline
$\delta_{8}^{D}-\delta_{8}$ & $25%
{{}^\circ}%
$ & $21%
{{}^\circ}%
$ & $20%
{{}^\circ}%
$\\\hline
\end{tabular}
\\
$\text{\textbf{Table\thinspace III:} }T,P_{t-c}$, on-shell $c\bar{c}$.
\end{tabular}
\]
Note that the fitting procedure is delicate because, to a large extend,
$P_{t-c}$ and $\chi T_{D}$ compete against each other (see Eq.(\ref{Eq24}))
and the $\chi_{\min}^{2}$ function is rather flat. The above results are those
for which both $\chi$ and $\delta_{8}^{D}-\delta_{8}$ are simultaneously small.

The $B\rightarrow\pi K$ branchings are now dominated by both $P_{t-c}$ and
$B\rightarrow\left\{  \bar{D}D_{s}\right\}  \rightarrow\pi K$ contributions,
and since these effects behave like a ''hadronic'' penguin, see Eq.(\ref{Eq24}%
), the pattern Eq.(\ref{Ex2}) is not altered. Concerning CP-asymmetries, we
find the pattern%
\begin{equation}
A_{\pi^{0}K^{+}}:A_{\pi^{+}K^{0}}:A_{\pi^{0}K^{0}}:A_{\pi^{-}K^{+}}%
\approx1:-\frac{P_{t-c}}{T}:-\frac{P_{t-c}}{T}:1 \label{Ex9}%
\end{equation}
Interestingly, some ratios of CP-asymmetries directly give the value of
$P_{t-c}/T$. These asymmetries are generated through the interference of
$T,P_{t-c}$ with $B\rightarrow\left\{  \bar{D}D_{s}\right\}  \rightarrow\pi
K$, and require only one strong phase $\delta_{8}^{D}-\delta_{8}$ of less than
about $25%
{{}^\circ}%
$ (generated by the mismatch between $PP\rightleftharpoons PP$ and $D\bar
{D}\rightleftharpoons D\bar{D}$, see Eq.(\ref{Eq18})).

In the $B\rightarrow\pi\pi$ sector, the fit is not very good (the large
$\chi_{\min}^{2}$ is essentially generated by the $\pi^{0}\pi^{0}$ and
$\pi^{0}\pi^{+}$ modes). Indeed, the $B\rightarrow\left\{  \bar{D}D\right\}
\rightarrow\pi\pi$ effects are Cabibbo-suppressed, and since there is no
averaging from $PP\rightleftharpoons PP$ strong phases, the decay branchings
keep their bare scalings in terms of dominant $T$ (see appendix B)%
\begin{equation}
Br\left(  \pi^{+}\pi^{0}\right)  :Br\left(  \pi^{+}\pi^{-}\right)  :Br\left(
\pi^{0}\pi^{0}\right)  \approx1/2:1:0 \label{Ex10}%
\end{equation}
which is rather far from current measurements (but not yet ruled out, see
table V, appendix C). Typically, $Br\left(  B\rightarrow\pi^{0}\pi^{0}\right)
<10^{-6}$ without $PP\rightleftharpoons PP$ strong phases (and assuming
$C/T\sim\mathcal{O}\left(  \lambda\right)  $ or smaller). Finally, $A_{\pi
^{+}\pi^{-}}$ is found a bit smaller (in the range 0.2 to 0.4), while
$S_{\pi^{+}\pi^{-}}$ behave roughly as in the SU(3)-elastic case (see Fig.5).%
\[%
\begin{tabular}
[c]{c}%
$%
{\includegraphics[
height=1.0603in,
width=5.6991in
]%
{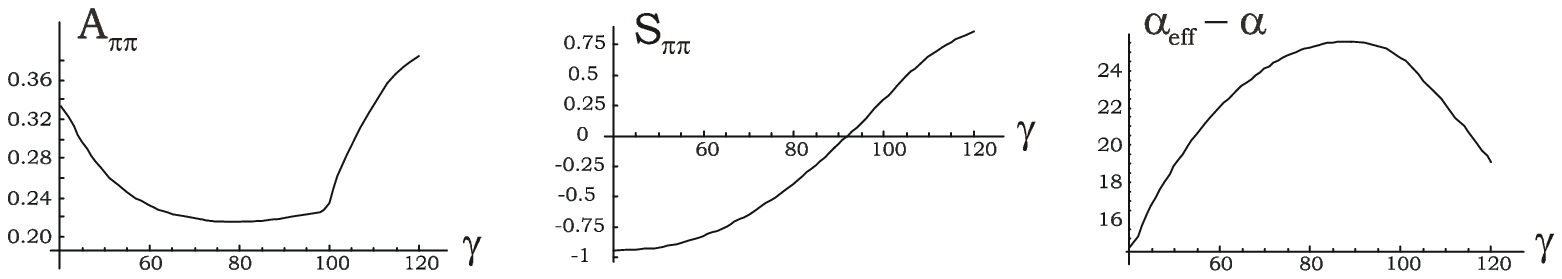}%
}%
$\\
\textbf{Figure 5:} Fit result for $A_{\pi^{+}\pi^{-}},S_{\pi^{+}\pi^{-}}$ and
$\alpha_{eff}$ under on-shell $c\bar{c}$ rescattering.
\end{tabular}
\ \
\]
Note, however, that $\alpha_{eff}\neq\alpha$ is now due to both $B\rightarrow
\left\{  D\bar{D}\right\}  \rightarrow\pi\pi$ and $P_{t-c}$, and therefore
that no $P_{t-c}$ penguin does not imply $\alpha_{eff}=\alpha$ (this is
obvious from Eq.(\ref{Eq24})).

For $B\rightarrow K\bar{K}$, the branchings are similar than in the
SU(3)-elastic case except for $B^{+}\rightarrow K^{+}K^{-}$ which is exactly
zero (no rescattering from $B\rightarrow\left\{  \pi^{+}\pi^{-}\right\}  $ and
no $B\rightarrow\left\{  D\bar{D}\right\}  \rightarrow K^{+}K^{-}$ because of
Eq.(\ref{Eq27})). The $B^{+}\rightarrow K^{+}\bar{K}^{0}$ and $B^{0}%
\rightarrow K^{0}\bar{K}^{0}$ direct CP-asymmetries tend to be much smaller
(between $0$ and $-0.3$), and in addition verifies%
\begin{equation}
A_{K^{+}\bar{K}^{0}}:A_{K^{0}\bar{K}^{0}}:A_{\pi^{0}\pi^{0}}=1:1:1
\label{Ex11}%
\end{equation}

If we now include the subleading $C$ and $P_{t-c}^{EW}$, the best fit
procedure produces a large $C$ amplitude to match the observed pattern of
$B\rightarrow\pi\pi$ branchings (we have restricted $C/T$ to be at most
$1/2$)
\[%
\begin{tabular}
[c]{c}%
\begin{tabular}
[c]{|lccc|}\hline
$\gamma$ & $60%
{{}^\circ}%
$ & $80%
{{}^\circ}%
$ & $100%
{{}^\circ}%
$\\\hline\hline
$\chi_{\min}^{2}$ & $18$ & $14$ & $12$\\\hline
$T$ & $0.49$ & $0.55$ & $0.62$\\
$C$ & $0.24$ & $0.20$ & $0.23$\\
$P_{t-c}$ & $0.09$ & $0.09$ & $0.09$\\
$P_{t-c}^{EW}$ & $0.011$ & $0.006$ & $0.004$\\\hline
$\chi$ & $3.5%
{{}^\circ}%
$ & $2.6%
{{}^\circ}%
$ & $4.7%
{{}^\circ}%
$\\\hline
$\delta_{8}^{D}-\delta_{8}$ & $26%
{{}^\circ}%
$ & $26%
{{}^\circ}%
$ & $15%
{{}^\circ}%
$\\\hline
\end{tabular}
\\
$\text{\textbf{Table\thinspace IV:} }T,P_{t-c},C,P_{t-c}^{EW}$, on-shell
$c\bar{c}$.
\end{tabular}
\]
The pattern of $B\rightarrow\pi K$ direct CP-asymmetries is much affected by
such a large $C$ amplitude%
\begin{equation}
A_{\pi^{0}K^{+}}:A_{\pi^{+}K^{0}}:A_{\pi^{0}K^{0}}:A_{\pi^{-}K^{+}}%
=1+\frac{C}{T}+\frac{CP_{t-c}}{T^{2}}:-\frac{P_{t-c}}{T}:-\frac{P_{t-c}+C}%
{T}-\frac{CP_{t-c}}{T^{2}}:1 \label{Ex12}%
\end{equation}
while corrections due to $P_{t-c}^{EW}$ are subleading. Notice, however, that
for the range of values in Table IV, this pattern is still different from the
SU(3)-elastic one Eq.(\ref{Ex3}). Finally, the pattern Eq.(\ref{Ex11}) survive
only partially: while $A_{K^{+}\bar{K}^{0}}$ and $A_{K^{0}\bar{K}^{0}}$ remain
exactly equal (no $C$ or $P_{t-c}^{EW}$ contribution), $A_{\pi^{0}\pi^{0}}$ is
dominated by $C$ and varies much.

To close this section, let us make a few comments. First, note that
phenomenologically speaking, $P_{t-c}/T=0$ is not yet ruled out by experiment
once the penguin-like rescattering effects $B\rightarrow\left\{  D\bar
{D}\right\}  \rightarrow PP$ are allowed. Indeed, the $B\rightarrow K\pi$
branchings could be dominated by the $B\rightarrow\left\{  D_{s}\bar
{D}\right\}  \rightarrow\pi K$ contributions, while both $B\rightarrow\pi K$
and $B\rightarrow\pi^{+}\pi^{-}$ direct CP-asymmetries could be generated by
interference of $B\rightarrow\left\{  D\bar{D}\right\}  \rightarrow PP$
contributions with $T$. In other words, assuming that $D\bar{D}%
\rightleftharpoons PP$ is the dominant rescattering process, the current data
requires $\left|  P_{t-c}^{eff}/T\right|  \neq0$ with $P_{t-c}^{eff}$ given in
Eq.(\ref{Eq24}), but do not say anything yet about the relative strength of
the quark-level penguin $P_{t-c}$ vs. its hadronic-level counterpart
$B\rightarrow\left\{  D\bar{D}\right\}  \rightarrow PP$. Now theoretically,
from short-distance analyses there is good reasons to expect $P_{t-c}/T\neq0$,
and we hope that the phenomenological patterns of CP-asymmetries described
above will help in extracting it from the data.

Second, under exact flavor SU(3), it appears that consistency with
$B\rightarrow\pi K$ direct CP-asymmetry measurements restricts $A_{\pi\pi}$ to
be smaller than $\sim0.5$. This statement is independent of the precise
physics in the charmed meson sector. Except for the values quoted for $\chi$,
the results of the fits are equally valid when the intermediate charmed meson
state is $D\bar{D}$, $D^{\ast}\bar{D}^{\ast},...$ or a charmonium state like
$P\eta_{c}$ (see \cite{Barshay}), or also combinations of them. Further, the
rescatterings between these states is irrelevant for $B\rightarrow PP$. Note
also that abandoning the coherence requirement Eq.(\ref{Eq27}) does not lead
to larger $A_{\pi\pi}$, even if $PA_{c}^{eff}=PA_{c}+2/3\beta+\alpha
\approx2/3\beta+\alpha$ (see Eq.(\ref{Eq20})) contributes to $B^{0}%
\rightarrow\pi^{+}\pi^{-}$ and not to $B\rightarrow\pi K$, because
$PA_{c}^{eff}$ also contributes to $B^{0}\rightarrow K^{+}K^{-}$ and therefore
cannot be large.

Finally, it should be noted that many relations between $B^{0},B^{+}$ direct
CP-asymmetry and $B_{S}$ ones exist
\[%
\begin{tabular}
[c]{ll}%
$A_{B_{S}\rightarrow\pi^{+}K^{-}}=A_{B^{0}\rightarrow\pi^{+}\pi^{-}}$ &
$A_{B_{S}\rightarrow K^{0}\bar{K}^{0}}=A_{B^{+}\rightarrow\pi^{+}K^{0}}$\\
$A_{B_{S}\rightarrow\pi^{0}\bar{K}^{0}}=A_{B^{0}\rightarrow\pi^{0}\pi^{0}%
}\;\;\;\;\;\;$ & $A_{B_{S}\rightarrow K^{+}K^{-}}=A_{B^{0}\rightarrow\pi
^{-}K^{+}}$%
\end{tabular}
\]
All these relations are unaffected by $C$ or $P_{t-c}^{EW}$ contributions and
specific to the dominant on-shell $c\bar{c}$ rescattering source (no
$PP\rightleftharpoons PP$ rescatterings).

\section{Conclusions}

The $B$ to two charmless pseudoscalar decay processes are analyzed assuming
exact SU(3) for the amplitudes and rescatterings. The basic tool is Watson's
theorem for hadronic final state interactions, which allows for a
factorization of rescattering effects from bare decay amplitudes. In addition,
the various FSI processes can also be factorized and treated separately. In
the present work, we assume SU(3)-elasticity for the $PP\rightleftharpoons PP$
part of the rescatterings, while $D\bar{D}\rightleftharpoons PP$ elastic
rescatterings are shown to amount to the redefinition of the penguin amplitude
as%
\begin{equation}
\lambda_{t}P_{t-c}^{eff}=\lambda_{t}P_{t-c}-\lambda_{c}\sqrt{\frac{3}{5}%
}\left(  1-e^{i(\delta_{8}^{D}-\delta_{8})}\right)  \chi T_{D} \label{CCL1}%
\end{equation}
where $T_{D}$ is the tree diagram contributing to $B\rightarrow D\bar{D}$ and
$\chi$ the small mixing parameter governing the $D\bar{D}$ pollution of
$B\rightarrow PP$ (this parameter is not fixed by SU(3), and must be small
because of Eq.(\ref{Eq4b})). Importantly, the redefinition Eq.(\ref{CCL1})
also apply when the intermediate charmed meson state is $D^{\ast}\bar{D}%
^{\ast},D^{\ast\ast}\bar{D}^{\ast\ast},P\eta_{c},...$, or any combination of
them, and therefore Eq.(\ref{CCL1}) can be seen as a hadronic representation
for on-shell intermediate $c\bar{c}$.

Also, it should be clear that once the various rescattering effects can be
treated separately, different theoretical tools can be combined. This is well
illustrated by Eq.(\ref{CCL1}): if one has a definite computation scheme for
$P_{t-c}^{eff}$ accounting for (elastic or inelastic) intermediate on-shell
$c\bar{c}$, but suspects non-negligible elastic long-distance
$PP\rightleftharpoons PP$ rescatterings, the present formalism permits the
combination of the two. In other words, it is perfectly consistent to account
for SU(N)-elastic long distance rescatterings by acting with the corresponding
$\sqrt{S_{SU(N)}}$ on effective quark diagram amplitudes already containing
various rescattering effects, like those produced by QCD-based approaches.

In our phenomenological approach, many amplitudes can be safely ignored and we
are left with the set of free parameters Eq.(\ref{Eq28}), with, in addition,
the weak angle $\gamma$. To further restrict the number of free parameters,
one rescattering source is assumed dominant: either SU(3)-elasticity
($PP\rightleftharpoons PP$) or on-shell $c\bar{c}$ ($D\bar{D}%
\rightleftharpoons PP$). The most prominent features of each fit are:

\begin{quote}
\textbf{SU(3)-elasticity:} The $B\rightarrow\pi K$ direct CP-asymmetries tend
to require large strong phases. This is nice for $B\rightarrow\pi\pi$ (see
Eq.(\ref{Ex4})) but may be problematic for $B\rightarrow KK$ (those modes are
predicted close to their present experimental upper bounds). Also, it should
be noted that the SU(3)-elastic pattern of $B\rightarrow\pi K$ direct
CP-asymmetries Eq.(\ref{Ex3}) is rather insensitive to subleading quark
diagram amplitudes ($C$ and $P_{t-c}^{EW}$), and is not in good agreement with
present measurements (though the discrepancy is not yet significant).

\textbf{On-shell }$c\bar{c}$\textbf{:} A single strong phase difference
($\delta_{8}^{D}-\delta_{8}$) of around $25%
{{}^\circ}%
$ is sufficient to produce the required asymmetries in the $\pi K$ and $\pi
\pi$ sectors. This strong phase difference is generated by the mismatch
between $PP\rightleftharpoons PP$ and $D\bar{D}\rightleftharpoons D\bar{D}$
elastic rescattering channels. Such a small strong phase is nice from the
point of view of Regge computations \cite{Regge} and from the naive
expectation for two particle flying apart with large momentum
\cite{MBInfinite}, but also with the SU(2)-elastic isospin analysis of
$B\rightarrow D\pi$ \cite{Dpi}. Compared with the SU(3)-elastic case, the fit
is not very good in the $B\rightarrow\pi\pi$ sector (no rescattering from
$B\rightarrow\pi^{+}\pi^{-}$ leads to the approximate pattern Eq.(\ref{Ex10}%
)), but better for $B\rightarrow\pi K$ direct CP-asymmetries (the pattern
Eq.(\ref{Ex9}) is favored by current measurements).
\end{quote}

Combining both sources of rescattering would certainly improve the quality of
the fit as more free parameters are available. For example, the difficulty of
the on-shell $c\bar{c}$ fit in the $B\rightarrow\pi\pi$ sector is reduced by
including a small amount of $PP\rightleftharpoons PP$ effects (like
$\delta_{27}-\delta_{8}\approx20%
{{}^\circ}%
$). More precise measurements are necessary to pursue the analysis in that direction.\newline 

{\Large Acknowledgements}: Many thanks are due to J.-M. G\'{e}rard, J.
Pestieau and J. Weyers for their helpful comments. This work was supported by
the Federal Office for Scientific, Technical and Cultural Affairs through the
Interuniversity Attraction Pole P5/27, and by the Institut Interuniversitaire
des Sciences Nucl\'{e}aires.

\appendix                                  

\section{Rescattering Matrix for $T_{3}=Y=0$ states}

This appendix presents the construction of the enlarged SU(3) rescattering
matrix for the $T_{3}=Y=0$ states. We start with%
\begin{align*}
\left(
\begin{array}
[c]{c}%
\left|  27,2,0,0\right\rangle \\
\left|  27,1,0,0\right\rangle \\
\left|  27,0,0,0\right\rangle \\
\left|  8,1,0,0\right\rangle \\
\left|  8,0,0,0\right\rangle \\
\left|  1,0,0,0\right\rangle \\
\left|  8_{D},1,0,0\right\rangle \\
\left|  8_{D},0,0,0\right\rangle \\
\left|  1_{D},0,0,0\right\rangle
\end{array}
\right)   &  =\underbrace{\left(
\begin{tabular}
[c]{ccccccccc}%
\multicolumn{6}{c}{} & $0$ & $0$ & $0$\\
\multicolumn{6}{c}{} & $0$ & $0$ & $0$\\
\multicolumn{6}{c}{$O_{SU\left(  3\right)  }^{PP}$} & $0$ & $0$ & $0$\\
\multicolumn{6}{c}{} & $0$ & $0$ & $0$\\
\multicolumn{6}{c}{} & $0$ & $0$ & $0$\\
\multicolumn{6}{c}{} & $0$ & $0$ & $0$\\
$0$ & $0$ & $0$ & $0$ & $0$ & $0$ & \multicolumn{3}{c}{}\\
$0$ & $0$ & $0$ & $0$ & $0$ & $0$ & \multicolumn{3}{c}{$O_{SU\left(  3\right)
}^{D\bar{D}}$}\\
$0$ & $0$ & $0$ & $0$ & $0$ & $0$ & \multicolumn{3}{c}{}%
\end{tabular}
\right)  }\left(
\begin{array}
[c]{c}%
\left\{  K^{+}K^{-}\right\} \\
\left\{  K^{0}\overline{K}^{0}\right\} \\
\left\{  \eta_{8}\eta_{8}\right\} \\
\left\{  \pi^{0}\eta_{8}\right\} \\
\left\{  \pi^{+}\pi^{-}\right\} \\
\left\{  \pi^{0}\pi^{0}\right\} \\
\left\{  D^{+}D^{-}\right\} \\
\left\{  D^{0}\bar{D}^{0}\right\} \\
\left\{  D_{s}^{+}D_{s}^{-}\right\}
\end{array}
\right) \\
&  \;\;\;\;\;\;\;\;\;\;\;\;\;\;\;\;O_{SU\left(  3\right)  }%
\end{align*}
where (entries are written according to $x=sign\left(  x\right)  \sqrt{\left|
x\right|  }$)%
\begin{align*}
O_{SU\left(  3\right)  }^{PP}  &  =\left(
\begin{array}
[c]{cccccc}%
0 & 0 & 0 & 0 & -1/3 & 2/3\\
-1/5 & 1/5 & 0 & 3/5 & 0 & 0\\
-3/20 & -3/20 & 27/40 & 0 & 1/60 & 1/120\\
3/10 & -3/10 & 0 & 2/5 & 0 & 0\\
-1/10 & -1/10 & -1/5 & 0 & 2/5 & 1/5\\
-1/4 & -1/4 & -1/8 & 0 & -1/4 & -1/8
\end{array}
\right) \\
O_{SU\left(  3\right)  }^{D\bar{D}}  &  =\left(
\begin{array}
[c]{ccc}%
1/2 & 1/2 & 0\\
-1/6 & 1/6 & 2/3\\
-1/3 & 1/3 & -1/3
\end{array}
\right)
\end{align*}
To introduce rescattering between $D\bar{D}$ and $PP$, it suffices to mix
$\left|  8,1,0,0\right\rangle $ with $\left|  8_{D},1,0,0\right\rangle $,
$\left|  8,0,0,0\right\rangle $ with $\left|  8_{D},0,0,0\right\rangle $ and
$\left|  1,0,0,0\right\rangle $ with $\left|  1_{D},0,0,0\right\rangle $ by
defining%
\[
O_{\chi}=\left(
\begin{tabular}
[c]{ccccccccc}%
$1$ & $0$ & $0$ & $0$ & $0$ & $0$ & $0$ & $0$ & $0$\\
$0$ & $1$ & $0$ & $0$ & $0$ & $0$ & $0$ & $0$ & $0$\\
$0$ & $0$ & $1$ & $0$ & $0$ & $0$ & $0$ & $0$ & $0$\\
$0$ & $0$ & $0$ & $\cos\chi_{8}$ & $0$ & $0$ & $\sin\chi_{8}$ & $0$ & $0$\\
$0$ & $0$ & $0$ & $0$ & $\cos\chi_{8}$ & $0$ & $0$ & $\sin\chi_{8}$ & $0$\\
$0$ & $0$ & $0$ & $0$ & $0$ & $\cos\chi_{1}$ & $0$ & $0$ & $\sin\chi_{1}$\\
$0$ & $0$ & $0$ & $-\sin\chi_{8}$ & $0$ & $0$ & $\cos\chi_{8}$ & $0$ & $0 $\\
$0$ & $0$ & $0$ & $0$ & $-\sin\chi_{8}$ & $0$ & $0$ & $\cos\chi_{8}$ & $0 $\\
$0$ & $0$ & $0$ & $0$ & $0$ & $-\sin\chi_{1}$ & $0$ & $0$ & $\cos\chi_{1} $%
\end{tabular}
\right)
\]
Finally, the rescattering matrix is found as $\sqrt{S_{\chi}}=O_{SU\left(
3\right)  }^{t}.O_{\chi}^{t}.\sqrt{S_{diag}}.O_{\chi}.O_{SU\left(  3\right)  }
$, with, when the $\chi_{i}$ are small,
\[
S_{diag}=diag\left(  e^{2i\delta_{27}},e^{2i\delta_{27}},e^{2i\delta_{27}%
},e^{2i\delta_{8}},e^{2i\delta_{8}},e^{2i\delta_{1}},e^{2i\delta_{8}^{D}%
},e^{2i\delta_{8}^{D}},e^{2i\delta_{1}^{D}}\right)
\]

\section{Decay Amplitude Expressions}

The $B\rightarrow PP$ decay amplitudes, under the approximations discussed in
the text, are%
\[%
\begin{tabular}
[c]{rl}%
$\mathcal{A}(B^{+}\rightarrow K^{+}\bar{K}^{0})=$ & $\frac{e^{i\delta_{8}%
}-e^{i\delta_{27}}}{5}\left(  T+C\right)  +e^{i\delta_{8}}P_{t-c}^{eff}$\\
$\mathcal{A}(B^{0}\rightarrow K^{+}K^{-})=$ & $\frac{5\left(  1-e^{i\delta
_{8}}\right)  +\left(  e^{i\delta_{8}}-e^{i\delta_{27}}\right)  }%
{20}T-\frac{5\left(  1-e^{i\delta_{8}}\right)  -3\left(  e^{i\delta_{8}%
}-e^{i\delta_{27}}\right)  }{60}C+\frac{2\left(  1-e^{i\delta_{8}}\right)
}{3}P_{t-c}^{eff}$\\
$\mathcal{A}(B^{0}\rightarrow K^{0}\bar{K}^{0})=$ & $\frac{5\left(
1-e^{i\delta_{8}}\right)  +\left(  e^{i\delta_{8}}-e^{i\delta_{27}}\right)
}{20}T-\frac{5\left(  1-e^{i\delta_{8}}\right)  -3\left(  e^{i\delta_{8}%
}-e^{i\delta_{27}}\right)  }{60}C+\frac{2+e^{i\delta_{8}}}{3}P_{t-c}%
^{eff}\medskip$\\
$\mathcal{A}(B^{+}\rightarrow\pi^{+}\pi^{0})=$ & $\frac{e^{i\delta_{27}}%
}{\sqrt{2}}\left(  T+C\right)  $\\
$\mathcal{A}(B^{0}\rightarrow\pi^{+}\pi^{-})=$ & $\frac{5+8e^{i\delta_{8}%
}+7e^{i\delta_{27}}}{20}T-\frac{5\left(  1-e^{i\delta_{8}}\right)  +21\left(
e^{i\delta_{8}}-e^{i\delta_{27}}\right)  }{60}C+\frac{2+e^{i\delta_{8}}}%
{3}P_{t-c}^{eff}$\\
$\mathcal{A}(B^{0}\rightarrow\pi^{0}\pi^{0})=$ & $\frac{5\left(
1-e^{i\delta_{8}}\right)  +13\left(  e^{i\delta_{8}}-e^{i\delta_{27}}\right)
}{20\sqrt{2}}T-\frac{5+16e^{i\delta_{8}}+39e^{i\delta_{27}}}{60\sqrt{2}%
}C+\frac{2+e^{i\delta_{8}}}{3\sqrt{2}}P_{t-c}^{eff}\medskip$\\
$\mathcal{A}(B^{+}\rightarrow K^{+}\pi^{0})=$ & $\frac{e^{i\delta_{8}%
}+4e^{i\delta_{27}}}{5\sqrt{2}}\left(  T+C\right)  +\frac{e^{i\delta_{8}}%
}{\sqrt{2}}P_{t-c}^{eff}$\\
$\mathcal{A}(B^{+}\rightarrow K^{0}\pi^{+})=$ & $\frac{e^{i\delta_{8}%
}-e^{i\delta_{27}}}{5}\left(  T+C\right)  +e^{i\delta_{8}}P_{t-c}^{eff}$\\
$\mathcal{A}(B^{0}\rightarrow K^{+}\pi^{-})=$ & $\frac{3e^{i\delta_{8}%
}+2e^{i\delta_{27}}}{5}T+\frac{2\left(  e^{i\delta_{27}}-e^{i\delta_{8}%
}\right)  }{5}C+e^{i\delta_{8}}P_{t-c}^{eff}$\\
$\mathcal{A}(B^{0}\rightarrow K^{0}\pi^{0})=$ & $\frac{3\left(  e^{i\delta
_{27}}-e^{i\delta_{8}}\right)  }{5\sqrt{2}}T+\frac{2e^{i\delta_{8}%
}+3e^{i\delta_{27}}}{5\sqrt{2}}C-\frac{e^{i\delta_{8}}}{\sqrt{2}}P_{t-c}%
^{eff}\medskip$\\
$\mathcal{A}\left(  B^{+}\rightarrow\pi^{+}\eta_{8}\right)  =$ &
$\frac{2e^{i\delta_{8}}+3e^{i\delta_{27}}}{5\sqrt{6}}\left(  T+C\right)
+\frac{2e^{i\delta_{8}}}{\sqrt{6}}P_{t-c}^{eff}$\\
$\mathcal{A}\left(  B^{0}\rightarrow\pi^{0}\eta_{8}\right)  =$ &
$-\frac{e^{i\delta_{8}}}{\sqrt{3}}P_{t-c}^{eff}\medskip$\\
$\mathcal{A}\left(  B^{+}\rightarrow K^{+}\eta_{8}\right)  =$ &
$-\frac{e^{i\delta_{8}}-6e^{i\delta_{27}}}{5\sqrt{6}}\left(  T+C\right)
-\frac{e^{i\delta_{8}}}{\sqrt{6}}P_{t-c}^{eff}$\\
$\mathcal{A}\left(  B^{0}\rightarrow K^{0}\eta_{8}\right)  =$ &
$\frac{3\left(  e^{i\delta_{27}}-e^{i\delta_{8}}\right)  }{5\sqrt{6}%
}T+\frac{2e^{i\delta_{8}}+3e^{i\delta_{27}}}{5\sqrt{6}}C-\frac{e^{i\delta_{8}%
}}{\sqrt{6}}P_{t-c}^{eff}\medskip$\\
$\mathcal{A}\left(  B^{0}\rightarrow\eta_{8}\eta_{8}\right)  =$ &
$\frac{5\left(  1-e^{i\delta_{8}}\right)  -3\left(  e^{i\delta_{8}}%
-e^{i\delta_{27}}\right)  }{20\sqrt{2}}T-\frac{5-16e^{i\delta_{8}}%
-9e^{i\delta_{27}}}{60\sqrt{2}}C+\frac{2-e^{i\delta_{8}}}{3\sqrt{2}}%
P_{t-c}^{eff}$%
\end{tabular}
\]
The amplitudes for the charge-conjugate modes are identical, except for the
understood CKM coefficients. Decay amplitudes for $B_{S}$ mode are:%
\[%
\begin{tabular}
[c]{rl}%
$\mathcal{A}\left(  B_{S}\rightarrow K^{-}\pi^{+}\right)  =$ &
$\frac{3e^{i\delta_{8}}+2e^{i\delta_{27}}}{5}T-\frac{2\left(  e^{i\delta_{8}%
}-e^{i\delta_{27}}\right)  }{5}C+e^{i\delta_{8}}P_{t-c}^{eff}$\\
$\mathcal{A}\left(  B_{S}\rightarrow\bar{K}^{0}\pi^{0}\right)  =$ &
$-\frac{3\left(  e^{i\delta_{8}}-e^{i\delta_{27}}\right)  }{5\sqrt{2}%
}T+\frac{2e^{i\delta_{8}}+3e^{i\delta_{27}}}{5\sqrt{2}}C-\frac{e^{i\delta_{8}%
}}{\sqrt{2}}P_{t-c}^{eff}\medskip$\\
$\mathcal{A}\left(  B_{S}\rightarrow\bar{K}^{0}\eta_{8}\right)  =$ &
$-\frac{3\left(  e^{i\delta_{8}}-e^{i\delta_{27}}\right)  }{5\sqrt{6}%
}T+\frac{2e^{i\delta_{8}}+3e^{i\delta_{27}}}{5\sqrt{6}}C-\frac{e^{i\delta_{8}%
}}{\sqrt{6}}P_{t-c}^{eff}\medskip$\\
$\mathcal{A}(B_{S}\rightarrow K^{+}K^{-})=$ & $\frac{5+8e^{i\delta_{8}%
}+7e^{i\delta_{27}}}{20}T-\frac{5\left(  1-e^{i\delta_{8}}\right)  +21\left(
e^{i\delta_{8}}-e^{i\delta_{27}}\right)  }{60}C+\frac{2+e^{i\delta_{8}}}%
{3}P_{t-c}^{eff}$\\
$\mathcal{A}(B_{S}\rightarrow K^{0}\bar{K}^{0})=$ & $\frac{5\left(
1-e^{i\delta_{8}}\right)  +\left(  e^{i\delta_{8}}-e^{i\delta_{27}}\right)
}{20}T-\frac{5\left(  1-e^{i\delta_{8}}\right)  -3\left(  e^{i\delta_{8}%
}-e^{i\delta_{27}}\right)  }{60}C+\frac{2+e^{i\delta_{8}}}{3}P_{t-c}%
^{eff}\medskip$\\
$\mathcal{A}(B_{S}\rightarrow\pi^{+}\pi^{-})=$ & $\frac{5\left(
1-e^{i\delta_{8}}\right)  +\left(  e^{i\delta_{8}}-e^{i\delta_{27}}\right)
}{20}T-\frac{5\left(  1-e^{i\delta_{8}}\right)  -3\left(  e^{i\delta_{8}%
}-e^{i\delta_{27}}\right)  }{60}C+\frac{2\left(  1-e^{i\delta_{8}}\right)
}{3}P_{t-c}^{eff}$\\
$\mathcal{A}(B_{S}\rightarrow\pi^{0}\pi^{0})=$ & $\frac{5\left(
1-e^{i\delta_{8}}\right)  +\left(  e^{i\delta_{8}}-e^{i\delta_{27}}\right)
}{20\sqrt{2}}T-\frac{5\left(  1-e^{i\delta_{8}}\right)  -3\left(
e^{i\delta_{8}}-e^{i\delta_{27}}\right)  }{60\sqrt{2}}C+\frac{2\left(
1-e^{i\delta_{8}}\right)  }{3\sqrt{2}}P_{t-c}^{eff}\medskip$\\
$\mathcal{A}\left(  B_{S}\rightarrow\pi^{0}\eta_{8}\right)  =$ &
$\frac{3\left(  e^{i\delta_{8}}-e^{i\delta_{27}}\right)  }{5\sqrt{3}%
}T-\frac{2e^{i\delta_{8}}+3e^{i\delta_{27}}}{5\sqrt{3}}C\medskip$\\
$\mathcal{A}\left(  B_{S}\rightarrow\eta_{8}\eta_{8}\right)  =$ &
$\frac{5\left(  1-e^{i\delta_{8}}\right)  +9\left(  e^{i\delta_{8}}%
-e^{i\delta_{27}}\right)  }{20\sqrt{2}}T-\frac{5+8e^{i\delta_{8}}%
+27e^{i\delta_{27}}}{60\sqrt{2}}C+\frac{2\left(  1+e^{i\delta_{8}}\right)
}{3\sqrt{2}}P_{t-c}^{eff}$%
\end{tabular}
\]
Finally, assuming negligible $PP,D\bar{D}\rightleftharpoons D^{\ast}\bar
{D}^{\ast},D^{\ast\ast}\bar{D}^{\ast\ast},...$, the $B\rightarrow D\bar{D}$
decay amplitudes are%
\[%
\begin{tabular}
[c]{rl}%
$\mathcal{A}\left(  B^{+}\rightarrow D_{s}^{+}\bar{D}^{0}\right)
=\mathcal{A}\left(  B^{0}\rightarrow D_{s}^{+}D^{-}\right)  =$ &
$e^{i\delta_{8}^{D}}T_{D} $\\
$\mathcal{A}\left(  B^{+}\rightarrow D^{+}\bar{D}^{0}\right)  =\mathcal{A}%
\left(  B_{S}\rightarrow D^{+}D_{s}^{-}\right)  =$ & $e^{i\delta_{8}^{D}}T_{D}
$\\
$\mathcal{A}\left(  B^{0}\rightarrow D^{+}D^{-}\right)  =$ & $\frac{1}%
{3}e^{i\delta_{8}^{D}}\left(  2+e^{-i\delta_{8}}\right)  T_{D}$\\
$\mathcal{A}\left(  B^{0}\rightarrow D^{0}\bar{D}^{0}\right)  =-\mathcal{A}%
\left(  B^{0}\rightarrow D_{s}^{+}D_{s}^{-}\right)  =$ & $\frac{1}%
{3}e^{i\delta_{8}^{D}}\left(  1-e^{-i\delta_{8}}\right)  T_{D}$\\
$\mathcal{A}\left(  B_{S}^{0}\rightarrow D_{s}^{+}D_{s}^{-}\right)  =$ &
$\frac{1}{3}e^{i\delta_{8}^{D}}\left(  2+e^{-i\delta_{8}}\right)  T_{D}$\\
$\mathcal{A}\left(  B_{S}\rightarrow D^{0}\bar{D}^{0}\right)  =-\mathcal{A}%
\left(  B_{S}\rightarrow D^{+}D^{-}\right)  =$ & $\frac{1}{3}e^{i\delta
_{8}^{D}}\left(  1-e^{-i\delta_{8}}\right)  T_{D}$%
\end{tabular}
\]

\section{Fit Results for Branchings and Asymmetries}

The experimental data on $B$ to two pseudoscalar decays we shall use are
summarized in table V and VI, where direct CP-asymmetries are defined
according to the sign convention%
\begin{equation}
A_{CP}=\frac{\Gamma\left(  \bar{B}\rightarrow\bar{f}\right)  -\Gamma\left(
B\rightarrow f\right)  }{\Gamma\left(  \bar{B}\rightarrow\bar{f}\right)
+\Gamma\left(  B\rightarrow f\right)  } \label{Eq30}%
\end{equation}
Average is made over CLEO \cite{CLEO}, Belle \cite{Belle} and BaBar
\cite{BaBar} measurements (see also \cite{Fry}), assuming no correlation. The
implications of SU(3) symmetry bear on the decay amplitudes, so predictions
for branchings must be corrected to account for lifetime differences and
available phase-space. From \cite{PDG2002}, the lifetime correction factor is
$\Gamma_{tot}\left(  B^{+}\right)  /\Gamma_{tot}\left(  B^{0}\right)  =0.92$,
while that of phase space are of at most a few percents, and are neglected.
The branchings under brackets in table V are branching fractions for the
physical $\eta$ state and not for $\eta_{8}$, so that they are only
indicative. Indeed, the small admixture of singlet $\eta_{0}$ into the
physical $\eta$ state could lead to large effects. The last observables of
interest to us are the time-dependent asymmetry parameters%
\begin{equation}
A_{f\bar{f}}=\frac{\left|  \lambda_{f\bar{f}}\right|  ^{2}-1}{\left|
\lambda_{f\bar{f}}\right|  ^{2}+1}=A_{CP},\;\;\;S_{f\bar{f}}%
=\frac{2\operatorname{Im}\lambda_{f\bar{f}}}{\left|  \lambda_{f\bar{f}%
}\right|  ^{2}+1},\;\;\;\;\text{with }\lambda_{f\bar{f}}=e^{-2i\beta
}\frac{A\left(  \bar{B}\rightarrow f\bar{f}\right)  }{A\left(  B\rightarrow
f\bar{f}\right)  } \label{Eq31}%
\end{equation}
The experimental situation and averages are (with inflated errors, see
\cite{PDG2002}):%
\[%
\begin{tabular}
[c]{cccc}
& Belle $\text{\cite{BelleTDAs}}$ & $\text{BaBar \cite{BaBarTDAs}}$ &
$\text{Average}$\\\hline
$A_{\pi^{+}\pi^{-}}$ & $0.77\pm0.27\pm0.08$ & $0.19\pm0.19\pm0.05$ &
$0.38\pm0.27$\\
$S_{\pi^{+}\pi^{-}}$ & $-1.23\pm0.41\pm0.08$ & $-0.40\pm0.22\pm0.03$ &
$-0.58\pm0.34$%
\end{tabular}
\]
We have used only $A_{\pi^{+}\pi^{-}}=-C_{\pi^{+}\pi^{-}}$ as input.

The function we minimize to find the best-fit values is as usual%
\begin{equation}
\chi^{2}=\sum_{f}\left(  \dfrac{Br^{th}-Br^{\exp}}{\sigma_{Br}^{\exp}}\right)
^{2}+\sum_{f}\left(  \dfrac{A_{CP}^{th}-A_{CP}^{\exp}}{\sigma_{A_{CP}}^{\exp}%
}\right)  ^{2} \label{Eq33}%
\end{equation}
where the sum runs over measured decay branchings and asymmetries (12 inputs).
Upper bounds are implemented using the arc-tangent representation of the step
function (3 inputs).

Finally, the Wolfenstein parametrization \cite{Wolf} of the CKM matrix
elements is used%
\begin{equation}
V_{CKM}=\left(
\begin{array}
[c]{ccc}%
1-\frac{\lambda^{2}}{2}-\frac{\lambda^{4}}{8} & \lambda & A\lambda^{3}\left(
\rho-i\eta\right) \\
-\lambda & 1-\frac{\lambda^{2}}{2}-\left(  \frac{A^{2}}{2}+\frac{1}{8}\right)
\lambda^{4} & A\lambda^{2}\\
A\lambda^{3}\left(  1-\rho-i\eta\right)  & -A\lambda^{2}+A\lambda^{4}\left(
\frac{1}{2}-\rho-i\eta\right)  & 1-\frac{A^{2}\lambda^{4}}{2}%
\end{array}
\right)  \label{Eq33b}%
\end{equation}
and we take the central values of \cite{PDG2002}
\begin{equation}
\lambda=0.2196\pm0.0023,\;A=0.854\pm0.045,\;\sqrt{\rho^{2}+\eta^{2}}%
=0.43\pm0.07\ \label{Eq34}%
\end{equation}
and keep only the weak angle $\gamma=\arctan\eta/\rho$ as a free parameter (so
unitarity of $V_{CKM}$ is implied). For the parameter $S_{f\bar{f}}$, we use
the average $\sin2\beta=0.736\pm0.049$ extracted from charmonium modes
\cite{Sin2beta}.\newline 

The result for branchings and asymmetries for the fits are collected below.
Column labels refer to the best-fit parameter tables given in the text.\newline %

\small
\begin{tabular}
[c]{c}%
$%
\begin{tabular}
[c]{|c|c|c|cc|cc|cc|cc|}\hline
\multicolumn{3}{|c|}{Branching$\;(\times10^{-6})$} &
\multicolumn{4}{|c|}{SU(3)-elastic} & \multicolumn{4}{|c|}{On-shell c\={c}%
}\\\cline{4-11}%
\multicolumn{3}{|c|}{} & \multicolumn{2}{|c}{I} & \multicolumn{2}{|c|}{II} &
\multicolumn{2}{|c|}{III} & \multicolumn{2}{|c|}{IV}\\\hline\hline
\multicolumn{2}{|c|}{$\Delta S=0$} & Exp. & 60$%
{{}^\circ}%
$ & 90$%
{{}^\circ}%
$ & 60$%
{{}^\circ}%
$ & 90$%
{{}^\circ}%
$ & 60$%
{{}^\circ}%
$ & 90$%
{{}^\circ}%
$ & 60$%
{{}^\circ}%
$ & 90$%
{{}^\circ}%
$\\\hline\hline
& $\overline{K^{0}}\eta_{8}$ & $-$ & 0.43 & 0.37 & 0.31 & 0.42 & 0.13 & 0.18 &
0.22 & 0.45\\
$B_{S}$ & $\overline{K^{0}}\pi^{0}$ & $-$ & 1.29 & 1.10 & 0.93 & 1.26 & 0.38 &
0.53 & 0.65 & 1.35\\
& $K^{-}\pi^{+}$ & $-$ & 5.91 & 5.78 & 5.72 & 4.61 & 4.77 & 4.72 & 4.55 &
4.24\\\hline\hline
& $\pi^{+}\eta_{8}$ & [4.0$\pm$0.9] & 1.59 & 1.38 & 2.41 & 1.72 & 1.42 &
1.17 & 2.31 & 1.93\\
$B^{+}$ & $\overline{K^{0}}K^{+}$ &
$<$%
1.3 & 1.09 & 1.28 & 1.01 & 1.28 & 0.88 & 1.21 & 0.89 & 1.18\\
& $\pi^{+}\pi^{0}$ & 5.3$\pm$0.8 & 3.48 & 4.42 & 5.43 & 5.67 & 2.07 & 3.10 &
4.34 & 5.86\\\hline\hline
& $K^{+}K^{-}$ &
$<$%
0.6 & 0.60 & 0.60 & 0.60 & 0.15 & 0 & 0 & 0 & 0\\
& $K^{0}\overline{K^{0}}$ &
$<$%
1.6 & 1.02 & 1.31 & 0.93 & 1.14 & 0.81 & 1.11 & 0.82 & 1.08\\
$B^{0}$ & $\eta_{8}\eta_{8}$ & $-$ & 0.17 & 0.25 & 0.12 & 0.09 & 0.05 & 0.06 &
0.11 & 0.09\\
& $\pi^{0}\eta_{8}$ & $\left[  \text{%
$<$%
2.9}\right]  $ & 0.26 & 0.38 & 0.26 & 0.38 & 0.27 & 0.37 & 0.27 & 0.36\\
& $\pi^{+}\pi^{-}$ & 4.6$\pm$0.4 & 4.82 & 4.79 & 4.58 & 4.53 & 5.03 & 4.98 &
4.80 & 4.47\\
& $\pi^{0}\pi^{0}$ & 1.92$\pm$0.44 & 2.12 & 1.70 & 1.77 & 1.57 & 0.41 & 0.56 &
0.68 & 1.43\\\hline\hline
\multicolumn{2}{|c|}{$\Delta S=1$} & Exp. & 60$%
{{}^\circ}%
$ & 90$%
{{}^\circ}%
$ & 60$%
{{}^\circ}%
$ & 90$%
{{}^\circ}%
$ & 60$%
{{}^\circ}%
$ & 90$%
{{}^\circ}%
$ & 60$%
{{}^\circ}%
$ & 90$%
{{}^\circ}%
$\\\hline\hline
& $K^{0}\eta_{8}$ & [2.4$\pm$0.9] & 3.33 & 3.17 & 2.94 & 2.82 & 3.37 & 3.17 &
2.92 & 2.86\\
$B^{0}$ & $K^{0}\pi^{0}$ & 11.2$\pm$1.4 & 9.98 & 9.51 & 8.81 & 8.45 & 10.1 &
9.51 & 8.75 & 8.57\\
& $K^{+}\pi^{-}$ & 18.2$\pm$0.8 & 18.6 & 19.2 & 18.3 & 19.0 & 18.6 & 19.2 &
18.5 & 19.1\\\hline\hline
& $K^{+}\eta_{8}$ & [3.1$\pm$0.7] & 4.02 & 3.52 & 3.68 & 3.20 & 4.02 & 3.51 &
3.50 & 3.20\\
$B^{+}$ & $K^{+}\pi^{0}$ & 12.8$\pm$1.1 & 10.3 & 10.4 & 11.7 & 11.7 & 10.1 &
10.4 & 11.6 & 11.6\\
& $K^{0}\pi^{+}$ & 20.6$\pm$1.4 & 22.0 & 20.6 & 22.3 & 20.7 & 22.0 & 20.6 &
21.8 & 20.5\\\hline\hline
& $K^{+}K^{-}$ & $-$ & 15.5 & 14.7 & 14.2 & 16.9 & 17.6 & 18.2 & 17.6 & 18.1\\
& $K^{0}\overline{K^{0}}$ & $-$ & 16.6 & 14.5 & 15.9 & 17.0 & 19.2 & 18.0 &
19.0 & 17.9\\
$B_{S}$ & $\eta_{8}\eta_{8}$ & $-$ & 14.1 & 12.5 & 12.9 & 14.2 & 17.0 & 16.0 &
15.8 & 15.2\\
& $\pi^{0}\eta_{8}$ & $-$ & 0.03 & 0.02 & 0.05 & 0.04 & 0 & 0 & 0.05 & 0.03\\
& $\pi^{+}\pi^{-}$ & $-$ & 5.19 & 6.95 & 6.55 & 1.95 & 0 & 0 & 0 & 0\\
& $\pi^{0}\pi^{0}$ & $-$ & 2.59 & 3.48 & 3.28 & 0.97 & 0 & 0 & 0 & 0\\\hline
\end{tabular}
$\\
\textbf{Table V}: Fit results for charge-average branching fractions.
\end{tabular}%

\small
$%
\begin{tabular}
[c]{c}%
$%
\begin{tabular}
[c]{|c|c|c|cc|cc|cc|cc|}\hline
\multicolumn{3}{|c}{Direct CP-asymmetry} & \multicolumn{4}{|c}{SU(3)-elastic}
& \multicolumn{4}{|c|}{On-shell c\={c}}\\\cline{4-11}\cline{4-11}%
\multicolumn{3}{|c}{$A_{CP}\;\left(  \%\right)  $} & \multicolumn{2}{|c}{I} &
\multicolumn{2}{|c|}{II} & \multicolumn{2}{|c|}{III} &
\multicolumn{2}{|c|}{IV}\\\hline\hline
\multicolumn{2}{|c|}{$\Delta S=0$} & Exp. & 60$%
{{}^\circ}%
$ & 90$%
{{}^\circ}%
$ & 60$%
{{}^\circ}%
$ & 90$%
{{}^\circ}%
$ & 60$%
{{}^\circ}%
$ & 90$%
{{}^\circ}%
$ & 60$%
{{}^\circ}%
$ & 90$%
{{}^\circ}%
$\\\hline\hline
& $\overline{K^{0}}\eta_{8}$ & $-$ & -78 & -99 & -88 & -70 & -32 & -19 & -72 &
-36\\
$B_{S}$ & $\overline{K^{0}}\pi^{0}$ & $-$ & -78 & -99 & -88 & -70 & -32 &
-19 & -72 & -36\\
& $K^{-}\pi^{+}$ & $-$ & 23 & 25 & 19 & 26 & 23 & 22 & 26 & 31\\\hline\hline
& $\pi^{+}\eta_{8}$ & $-$ & 49 & 60 & 28 & 41 & 23 & 27 & 27 & 35\\
$B^{+}$ & $\overline{K^{0}}K^{+}$ & $-$ & -71 & -64 & -67 & -56 & -32 & -19 &
-33 & -24\\
& $\pi^{+}\pi^{0}$ & -7$\pm$15 & 0 & 0 & 0 & 0 & 0 & 0 & 0 & 0\\\hline\hline
& $K^{+}K^{-}$ & $-$ & 14 & 16 & 11 & 18 & - & - & - & -\\
& $K^{0}\overline{K^{0}}$ & $-$ & -88 & -95 & -87 & -58 & -32 & -19 & -33 &
-24\\
$B^{0}$ & $\eta_{8}\eta_{8}$ & $-$ & -96 & -98 & -16 & 39 & -32 & -19 & 22 &
33\\
& $\pi^{0}\eta_{8}$ & $-$ & 0 & 0 & 0 & 0 & -32 & -19 & -33 & -24\\
& $\pi^{+}\pi^{-}$ & 38$\pm$27 & 42 & 54 & 43 & 42 & 23 & 22 & 26 & 31\\
& $\pi^{0}\pi^{0}$ & $-$ & -49 & -70 & -69 & -85 & -32 & -19 & -72 &
-36\\\hline\hline
\multicolumn{2}{|c|}{$\Delta S=1$} & Exp. & 60$%
{{}^\circ}%
$ & 90$%
{{}^\circ}%
$ & 60$%
{{}^\circ}%
$ & 90$%
{{}^\circ}%
$ & 60$%
{{}^\circ}%
$ & 90$%
{{}^\circ}%
$ & 60$%
{{}^\circ}%
$ & 90$%
{{}^\circ}%
$\\\hline\hline
& $K^{0}\eta_{8}$ & $-$ & 10 & 12 & 9.5 & 11 & 1.2 & 1.1 & 5.5 & 5.9\\
$B^{0}$ & $K^{0}\pi^{0}$ & 3$\pm$37 & 10 & 12 & 9.5 & 11 & 1.2 & 1.1 & 5.5 &
5.9\\
& $K^{+}\pi^{-}$ & -9.3$\pm$2.9 & -7.5 & -7.7 & -6.1 & -6.4 & -6.1 & -5.5 &
-6.7 & -7.0\\\hline\hline
& $K^{+}\eta_{8}$ & $-$ & 19 & 23 & 18 & 22 & 7.4 & 7.6 & 13 & 15\\
$B^{+}$ & $K^{+}\pi^{0}$ & 1$\pm$12 & -15 & -15 & -11 & -12 & -6.1 & -5.5 &
-9.1 & -9.8\\
& $K^{0}\pi^{+}$ & 1$\pm$6 & 3.4 & 3.9 & 2.9 & 3.4 & 1.2 & 1.1 & 1.3 &
1.3\\\hline\hline
& $K^{+}K^{-}$ & $-$ & -12 & -16 & -13 & -11 & -6.1 & -5.5 & -6.7 & -7.0\\
& $K^{0}\overline{K^{0}}$ & $-$ & 5.0 & 7.9 & 4.7 & 3.6 & 1.2 & 1.1 & 1.3 &
1.3\\
$B_{S}$ & $\eta_{8}\eta_{8}$ & $-$ & 8.1 & 11 & 9.5 & 8.7 & 1.2 & 1.1 & 3.3 &
3.5\\
& $\pi^{0}\eta_{8}$ & $-$ & 0 & 0 & -80 & -71 & - & - & 0 & 0\\
& $\pi^{+}\pi^{-}$ & $-$ & -1.4 & -1.3 & -0.9 & -1.3 & - & - & - & -\\
& $\pi^{0}\pi^{0}$ & $-$ & -1.4 & -1.3 & -0.9 & -1.3 & - & - & - & -\\\hline
\end{tabular}
\ $\\
\textbf{Table VI}: Fit results for direct CP-asymmetries (A$_{CP}$).\\
\multicolumn{1}{l}{\ \ }%
\end{tabular}
\ \ $\newline %

\small
$%
\begin{tabular}
[c]{c}%
$%
\begin{tabular}
[c]{|c|c|c|cc|cc|cc|cc|}\hline
\multicolumn{3}{|c|}{$S_{f\bar{f}}$ $\left(  \%\right)  $} &
\multicolumn{4}{|c}{SU(3)-elastic} & \multicolumn{4}{|c|}{On-shell c\={c}%
}\\\cline{4-11}%
\multicolumn{3}{|c}{} & \multicolumn{2}{|c}{I} & \multicolumn{2}{|c|}{II} &
\multicolumn{2}{|c|}{III} & \multicolumn{2}{|c|}{IV}\\\hline\hline
\multicolumn{2}{|c}{$\Delta S=0$} & Exp. & 60$%
{{}^\circ}%
$ & 90$%
{{}^\circ}%
$ & 60$%
{{}^\circ}%
$ & 90$%
{{}^\circ}%
$ & 60$%
{{}^\circ}%
$ & 90$%
{{}^\circ}%
$ & 60$%
{{}^\circ}%
$ & 90$%
{{}^\circ}%
$\\\hline\hline
& $K^{+}K^{-}$ & $-$ & -99 & -83 & -99 & -90 & - & - & - & -\\
& $K^{0}\overline{K^{0}}$ & $-$ & -30 & 2.3 & -27 & -5.2 & -19 & -10 & -24 &
-16\\
$B^{0}$ & $\eta_{8}\eta_{8}$ & $-$ & -23 & -18 & -49 & -36 & -19 & -10 & -97 &
-86\\
& $\pi^{0}\eta_{8}$ & $-$ & 6.2 & -1.2 & 6.2 & -1.2 & -19 & -10 & -24 & -16\\
& $\pi^{+}\pi^{-}$ & -58$\pm$34 & -69 & 16 & -67 & -1.3 & -81 & -6.5 & -81 &
-9.7\\
& $\pi^{0}\pi^{0}$ & $-$ & -67 & 8.4 & -39 & 51 & -19 & -10 & 65 & 86\\\hline
\end{tabular}
$\\
\textbf{Table VII}: Fit results for the time-dependent CP-asymmetry parameter
$S_{f\bar{f}}$.
\end{tabular}
$
\end{document}